\documentclass[letterpaper]{article}
\usepackage[preprint]{aaai2027}
\usepackage[hyphens]{url}
\usepackage{graphicx}
\usepackage{amsmath}
\usepackage{booktabs}
\usepackage{tabularx}
\usepackage{array}
\usepackage{colortbl}
\usepackage{multirow}
\usepackage{tikz}
\usepackage{placeins}
\usepackage{float}
\usepackage{listings}
\usetikzlibrary{arrows.meta,backgrounds,calc,fit,positioning,shapes.geometric}
\usepackage{natbib}
\newcolumntype{Y}{>{\raggedright\arraybackslash}X}
\newcolumntype{C}{>{\centering\arraybackslash}X}
\newcolumntype{L}[1]{>{\raggedright\arraybackslash}p{#1}}
\definecolor{feedbacksuccess}{RGB}{218,237,218}
\definecolor{feedbackfailure}{RGB}{255,229,204}
\definecolor{feedbacknormal}{RGB}{207,226,243}
\definecolor{positivecell}{RGB}{225,239,237}
\definecolor{negativecell}{RGB}{248,233,222}
\definecolor{missingcell}{RGB}{241,241,241}
\definecolor{selectedskill}{RGB}{232,242,247}
\definecolor{positiveink}{RGB}{0,105,92}
\definecolor{negativeink}{RGB}{178,78,26}
\definecolor{neutralink}{RGB}{100,100,100}
\newcommand{\selcell}[1]{\cellcolor{selectedskill}\bfseries #1}
\newcommand{\poscell}[1]{\cellcolor{positivecell}#1}
\newcommand{\negcell}[1]{\cellcolor{negativecell}#1}
\newcommand{\zerocell}[1]{\cellcolor{missingcell}#1}
\newcommand{\failcell}[1]{\cellcolor{missingcell}#1}

\newcommand{\scorebase}[1]{#1 (\textcolor{neutralink}{--})}
\newcommand{\scorepos}[2]{#1 (\textcolor{positiveink}{#2})}
\lstdefinestyle{skillartifact}{
  basicstyle=\ttfamily\tiny,
  breaklines=true,
  breakatwhitespace=false,
  columns=fullflexible,
  keepspaces=true,
  showstringspaces=false,
  frame=single,
  framerule=0.25pt,
  xleftmargin=0.4em,
  xrightmargin=0.4em,
  aboveskip=0.4em,
  belowskip=0.2em
}
\newcommand{\scoreneg}[2]{#1 (\textcolor{negativeink}{#2})}
\newcommand{\scorezero}[2]{#1 (\textcolor{neutralink}{#2})}

\newcommand{\notavailablecell}{\cellcolor{missingcell}N/A}

\title{Rethinking Self-Evolving Agent Skills: Feedback Dynamics over Multiple Rounds}
\author{
Yuxuan Liu\textsuperscript{\rm 1},
Zhaochen Su\textsuperscript{\rm 1},
Yuhao Zhang\textsuperscript{\rm 3},
Jiahe Guo\textsuperscript{\rm 2},
Zhongwei Xie\textsuperscript{\rm 1},
Huihao Jing\textsuperscript{\rm 1},\\
Lingyun Xie\textsuperscript{\rm 1},
Qing Zong\textsuperscript{\rm 1},
Yauwai Yim\textsuperscript{\rm 1},
Zhixiong Zhang\textsuperscript{\rm 4},
Haoran Li\textsuperscript{\rm 1}\corresponding,
Yangqiu Song\textsuperscript{\rm 1}
}
\affiliations{
\textsuperscript{\rm 1}The Hong Kong University of Science and Technology \quad
\textsuperscript{\rm 2}Harbin Institute of Technology\\
\textsuperscript{\rm 3}Harbin Institute of Technology, Shenzhen \quad
\textsuperscript{\rm 4}Shanghai Jiao Tong University\\
\{yliurk,zsubf,hlibt\}@connect.ust.hk, yqsong@cse.ust.hk
}

\begin{document}

\maketitle

\begin{abstract}
Self-evolving skill systems promise to improve agents by turning execution feedback into persistent skill updates without changing the underlying model. Yet it remains unclear when further evolution helps, how successful and failed trajectories shape revision, and whether extra test-time computation can recover the same gains. To address these questions, we present a controlled evaluation framework across five benchmarks and three models. Our primary study contains 42 feedback runs across 14 supported model--benchmark settings. Within each setting, we hold the executor and optimizer configuration, revision procedure, validation rule, and round budget fixed, while varying only the feedback shown to the optimizer: successes and failures (Normal), failures only, or successes only. Evolution is sparse: only 55 of 388 candidates establish byte-distinct validation bests. Validation-based selection chooses an evolved skill in 11 of 14 settings, nine of which improve released-test performance. All 11 selections come from feedback conditions that include failed trajectories, although the relative ranking of Normal and Fail-only varies across settings. Validation and downstream evaluations on test, robustness, and transfer sometimes favor different feedback views. A broader SearchQA analysis covering eight models shows similarly sparse, feedback-dependent dynamics. In the GPT-5.5 test-time-scaling controls, oracle Parallel Sampling comes within 0.43 points of the evolved SearchQA skill but remains 30.96 points behind on SpreadsheetBench; Sequential Refinement recovers neither gain. Overall, persistent skill self-evolution is better understood as sparse, validation-filtered search with model- and benchmark-dependent returns, rather than steady improvement from additional rounds. The implementation is available at \url{https://github.com/HKUST-KnowComp/rethinkskill}.
\end{abstract}

\section{Introduction}

Agent skills encode reusable behavior outside the underlying model and guide later executions without updating model parameters. Self-evolving skill systems extend this idea by using execution traces and evaluation feedback to generate, repair, and retain new skill versions. Unlike extra inference for a single task instance, a revised skill persists across future executions. Prior work has reported substantial gains over skill-free baselines and, in some settings, human-authored skills \cite{li2026skillhone,liu2026skillrevise,yang2026skillopt}.

A central design choice is which execution evidence should shape the next skill version. Success-gated methods retain useful discoveries from successful trajectories \cite{liu2026skillsvote}. Diagnosis-oriented methods revise skills based on observed failures \cite{liu2026skillrevise,liu2026skillforge}. Mixed-trajectory methods use both sources \cite{yang2026skillopt,ni2026trace2skill,yang2026geoskill}. These choices reflect different intuitions: successes show behavior worth preserving, failures expose skill defects, and mixed feedback provides broader but potentially conflicting evidence.

However, existing results do not isolate the effect of feedback composition. Each system is usually evaluated under its own revision and selection procedure, leaving three questions unresolved. First, before-and-after scores hide accepted revisions, rejected candidates, rollbacks, and stopping decisions. They therefore do not show when additional rounds help or when the search has stopped making progress. Second, comparisons across systems cannot determine how successful trajectories, failed trajectories, or both affect evolution under the same procedure and across different models. Third, it remains unclear whether additional per-instance inference can achieve similar gains without changing the persistent skill. Run-to-run variation and verifier choice further complicate these comparisons because score changes do not always reflect lasting skill changes.

To address these questions, we present a controlled evaluation framework for feedback-conditioned skill evolution. Our primary study contains 42 feedback runs across 14 supported model--benchmark settings involving GPT-5.5, Gemini 3.1 Pro, DeepSeek V4-Pro, and five benchmarks. Within each setting, Normal, Fail-only, and Success-only start from the same parent skill. The task executor and revision optimizer use the same model configuration, while the revision procedure, validation rule, and round budget remain fixed. At each round, the revision operator proposes a candidate skill. A candidate becomes the next-round skill when validation does not decrease; otherwise, the current skill is retained. Only strict improvement updates the best checkpoint.

For artifact-level analysis, we count a new validation best only when the candidate is also byte-distinct from the incoming skill. We record all candidates, acceptance and rollback decisions, best-so-far updates, and skill identities, exposing each view's full search trajectory. We retain the validation-best skill from each run and select one skill for each model--benchmark setting using validation alone; test and diagnostic results do not affect this selection. After evolution, we evaluate the frozen skills on released test, same-task robustness, and transfer. For the five GPT-5.5 settings, we also examine test-time scaling, repeated-deployment variability, and verifier sensitivity. A broader SearchQA analysis expands the model coverage to eight models in total. Figure~\ref{fig:lifecycle} summarizes the framework.

The results show that skill evolution is sparse and depends on both the model and benchmark. Only 55 of 388 candidates establish byte-distinct validation bests. The observed trajectories range from late discovery to early saturation and complete stagnation. Validation selects an evolved skill in 11 of 14 model--benchmark settings, nine of which improve released-test performance. Nine also improve robustness and nine improve transfer, but only seven improve both. All 11 selected evolved skills come from Normal or Fail-only: Normal is selected in nine settings and Fail-only in two, while Success-only is never selected in the primary study. The broader SearchQA analysis shows similar sparse, feedback-dependent dynamics across eight models. Test-time scaling is also uneven: oracle Parallel Sampling nearly recovers the SearchQA gain but remains far behind on SpreadsheetBench, while Sequential Refinement reproduces neither. Together, these results characterize persistent skill self-evolution as validation-filtered search rather than steady improvement from additional rounds.

This paper makes three contributions:
\begin{itemize}
    \item \textbf{Cross-model evolution dynamics.} We trace 388 candidates across 42 feedback runs and 14 model--benchmark settings, revealing late improvement, early saturation, rejected regressions, and stagnation. A broader eight-model SearchQA analysis shows similarly sparse, feedback-dependent dynamics across model families.

    \item \textbf{Controlled feedback comparison.} We compare Normal, Fail-only, and Success-only while holding the evolution procedure fixed. All 11 evolved selections in the primary study include failed trajectories, while artifact-level case studies show how different feedback views produce different retained guidance.

    \item \textbf{Evolution, generalization, and test-time computation.} We evaluate whether selected skills improve test, robustness, and transfer, and compare the GPT-5.5 skills with parallel and sequential test-time scaling. The results separate persistent skill gains from per-instance inference and validation-specific improvement.
\end{itemize}

\section{Related Work}

\subsection{Self-Evolving Agent Skills}

Self-evolving skill systems make different choices about which execution evidence becomes persistent skill change \cite{jiang2026xskill,zhang2026coevoskills,tian2026skillscoach,xia2026skillrl,shen2026skilloptlite}. Success-gated systems such as SkillsVote admit only successful reusable discoveries to evidence-gated skill-library updates \cite{liu2026skillsvote}, whereas diagnosis-oriented systems such as SkillRevise and SkillForge identify execution failures, localize skill defects, and revise the affected skills \cite{liu2026skillrevise,liu2026skillforge}. Mixed-trajectory systems, including SkillOpt, Trace2Skill, SkillGen, and OptSkills, consolidate successful procedures and failure-derived lessons into subsequent skill revisions \cite{yang2026skillopt,ni2026trace2skill,ma2026skillgen,yang2026optskills}. These systems evaluate their respective native feedback policies as a whole, rather than holding the update and validation procedure fixed while comparing successful trajectories only, failed trajectories only, and both together across complete multi-round evolution runs. Consequently, the comparative roles of successful and failed execution evidence in adaptive persistent-skill search remain unclear.

\paragraph{Agent skill evaluation and benchmarks.}
Agent-skill benchmarks evaluate marginal utility, continual learning, and process quality at several levels. SkillsBench and SWE-Skills-Bench use paired conditions to isolate skill utility, while SkillLearnBench evaluates skill quality, execution trajectories, and task outcomes \cite{li2026skillsbench,han2026sweskillsbench,zhong2026skilllearnbench}. OpenClawBench complements outcome-based evaluation by aligning task-oracle outcomes with localized process-anomaly evidence in real-world execution trajectories \cite{liu2026openclawbench}. Other work analyzes experience extraction and consumption, process-level skill use, or paired behavioral influence \cite{huang2026rawexperience,zhu2026skillcoach,zhou2026cta}. SkillsWild evaluates skill retrieval and use under realistic library conditions, while SkillJuror measures behavioral changes under controlled skill organizations \cite{liu2026skillswild,chen2026skilljuror}. SEA-Eval studies sequential cross-task evolution, and metric co-evolution examines improvement when the evaluator itself changes \cite{jiang2026seaeval,zhang2026grader}. These benchmarks characterize skill utility and evaluation scope, but they do not isolate how alternative feedback sources shape a fixed-model skill-evolution procedure over multiple rounds.

\begin{table}[!t]
\centering
\small
\setlength{\tabcolsep}{4pt}
\renewcommand{\arraystretch}{1.2}
\begin{tabularx}{\columnwidth}{@{}X@{}}
\toprule
\rowcolor{black!8}\textbf{Feedback exposure}\\
\midrule
\rowcolor{feedbacksuccess}\textbf{Success-only}\\
\addlinespace[1pt]
\textbf{SkillsVote}~{\footnotesize\cite{liu2026skillsvote}}\\
\midrule
\rowcolor{feedbackfailure}\textbf{Failure-driven}\\
\addlinespace[1pt]
\textbf{SkillRevise}~{\footnotesize\cite{liu2026skillrevise}}\\
\textbf{SkillForge}~{\footnotesize\cite{liu2026skillforge}}\\
\textbf{EvoSkill}~{\footnotesize\cite{alzubi2026evoskill}}\\
\textbf{MemSkill}~{\footnotesize\cite{zhang2026memskill}}\\
\textbf{SkillAdaptor}~{\footnotesize\cite{yu2026skilladaptor}}\\
\midrule
\rowcolor{feedbacknormal}\textbf{Normal}\\
\addlinespace[1pt]
\textbf{SkillOpt}~{\footnotesize\cite{yang2026skillopt}}\\
\textbf{Trace2Skill}~{\footnotesize\cite{ni2026trace2skill}}\\
\textbf{GeoSkill}~{\footnotesize\cite{yang2026geoskill}}\\
\textbf{SkillGen}~{\footnotesize\cite{ma2026skillgen}}\\
\textbf{OptSkills}~{\footnotesize\cite{yang2026optskills}}\\
\bottomrule
\end{tabularx}
\caption{Representative self-evolving skill systems grouped by feedback evidence.}
\label{tab:feedback-taxonomy}
\end{table}

\begin{figure*}[!t]
\centering
\includegraphics[width=\textwidth]{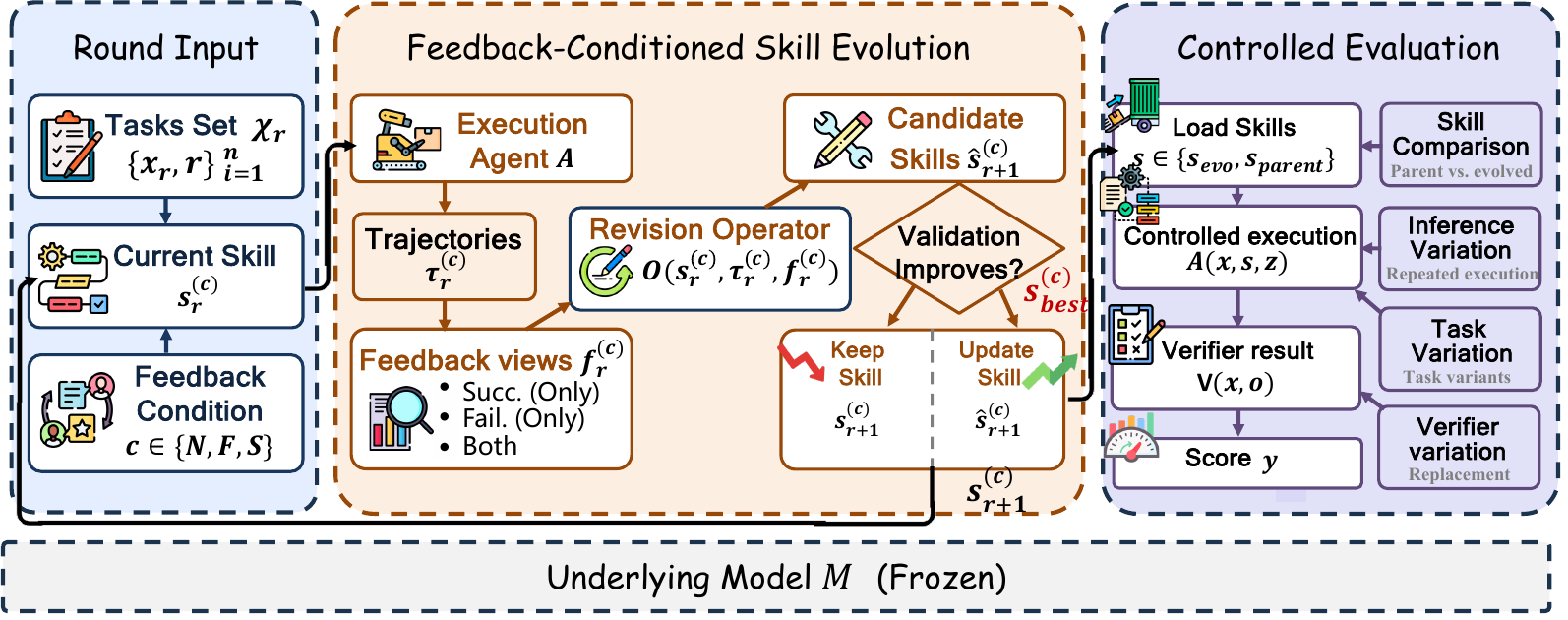}
\caption{Feedback-conditioned skill evolution and controlled evaluation. Each round executes the current skill, constructs a feedback view, and proposes a candidate for validation. The validation gate determines the next-round skill and updates the best checkpoint only after strict improvement. The best checkpoint is then frozen for controlled evaluation.}
\label{fig:lifecycle}
\end{figure*}

\section{Framework Overview}

\paragraph{Self-evolution process.}
Let $M$ denote the fixed underlying agent model and $s_r^{(c)}$ the persistent skill at evolution round $r$ under feedback condition $c$. Executing the round tasks with $M$ and $s_r^{(c)}$ produces trajectories $\tau_r^{(c)}$, while $f_r^{(c)}$ denotes the feedback exposed for revision. The same model instantiates the revision operator $O$, which proposes a candidate from the current skill and condition-specific evidence:
\begin{equation}
\widehat{s}_{r+1}^{(c)}=O(s_r^{(c)},\tau_r^{(c)},f_r^{(c)}).
\end{equation}
Here, $\widehat{s}_{r+1}^{(c)}$ is a candidate rather than the next-round skill. If the candidate is not rejected,
\begin{equation}
s_{r+1}^{(c)}\leftarrow\widehat{s}_{r+1}^{(c)}.
\end{equation}
Otherwise,
\begin{equation}
s_{r+1}^{(c)}\leftarrow s_r^{(c)}.
\end{equation}
We separately track the validation-best skill $s_{\mathrm{best}}^{(c)}$, initialized as $s_0^{(c)}$. For artifact-level analysis, a new best requires both a strict validation improvement and a byte-distinct candidate. Score changes from byte-identical reruns are treated as execution variability and do not update the reported best round. After the final round, each view returns its best validation skill. For each benchmark, we choose among these three skills using validation before evaluating released test, robustness, and transfer.

\paragraph{Framework components.}
Figure~\ref{fig:lifecycle} instantiates this process through three components.

\textit{Execution and feedback construction.}
The execution component applies the current skill to the round tasks and records the resulting trajectories, which the benchmark verifier labels as successful or failed. The feedback component then constructs one of three views: Normal includes both successful and failed trajectories, Fail-only includes failed trajectories, and Success-only includes successful trajectories.

\textit{Validation-gated skill revision.}
The revision component generates a candidate through $O$. The validation gate advances a candidate whose validation score does not decrease and carries the current skill forward after a regression. A separate checkpoint updates only after a strict validation improvement and is returned after evolution.

\textit{Controlled evaluation.}
After evolution, the resulting skills are frozen and passed through a common evaluation path. For a task $x$ and skill $s$, execution and verification are
\begin{equation}
o=A(x,s,z), \qquad y=V(x,o),
\end{equation}
where $z$ denotes the controlled execution conditions. Changes to $s$, $z$, $x$, and $V$ respectively organize the parent--evolved comparison, test-time scaling and repeated execution, task variation, and verifier replacement.

\raggedbottom
\section{Experiments}
\label{sec:experiments}

\begin{table*}[!t]
\centering
{\tiny
\setlength{\tabcolsep}{1.15pt}
\renewcommand{\arraystretch}{1.06}
\begin{tabularx}{\textwidth}{L{0.095\textwidth} L{0.082\textwidth} *{4}{C} @{\hspace{1.5pt}} *{4}{C} @{\hspace{1.5pt}} *{4}{C}}
\toprule
\textbf{Benchmark} & \textbf{Condition} &
\multicolumn{4}{c}{\textbf{GPT-5.5}} &
\multicolumn{4}{c}{\textbf{Gemini 3.1 Pro}} &
\multicolumn{4}{c}{\textbf{DeepSeek V4-Pro}} \\
\cmidrule(lr){3-6}\cmidrule(lr){7-10}\cmidrule(lr){11-14}
& & \textbf{Validation} & \textbf{Test} & \textbf{R} & \textbf{T}
& \textbf{Validation} & \textbf{Test} & \textbf{R} & \textbf{T}
& \textbf{Validation} & \textbf{Test} & \textbf{R} & \textbf{T} \\
\midrule
\multirow{4}{*}{\textbf{SearchQA}} & \mbox{Parent ($r=0$)}
& \scorebase{80.0} & \scorebase{75.6} & \scorebase{75.9} & \scorebase{72.1}
& \scorebase{79.0} & \scorebase{76.9} & \underline{\scorebase{76.3}} & \scorebase{73.7}
& \scorebase{74.5} & \scorebase{73.0} & \scorebase{72.4} & \scorebase{72.6} \\
& Normal
& \poscell{\selcell{\underline{\scorepos{82.0}{+2.0}}}} & \poscell{\selcell{\underline{\scorepos{77.9}{+2.3}}}} & \poscell{\selcell{\scorepos{76.1}{+0.2}}} & \poscell{\selcell{\underline{\scorepos{77.6}{+5.5}}}}
& \poscell{\selcell{\underline{\scorepos{82.0}{+3.0}}}} & \negcell{\selcell{\scoreneg{76.6}{-0.4}}} & \negcell{\selcell{\scoreneg{76.0}{-0.3}}} & \poscell{\selcell{\underline{\scorepos{75.2}{+1.5}}}}
& \poscell{\selcell{\underline{\scorepos{76.5}{+2.0}}}} & \poscell{\selcell{\scorepos{73.9}{+0.9}}} & \negcell{\selcell{\scoreneg{72.2}{-0.2}}} & \poscell{\selcell{\underline{\scorepos{74.4}{+1.8}}}} \\
& Fail-only
& \poscell{\scorepos{81.5}{+1.5}} & \poscell{\scorepos{77.3}{+1.7}} & \negcell{\scoreneg{74.9}{-1.0}} & \poscell{\scorepos{76.2}{+4.2}}
& \poscell{\scorepos{81.5}{+2.5}} & \poscell{\scorepos{77.6}{+0.6}} & \negcell{\scoreneg{76.0}{-0.3}} & \poscell{\scorepos{74.6}{+0.8}}
& \poscell{\scorepos{76.0}{+1.5}} & \poscell{\underline{\scorepos{74.2}{+1.2}}} & \poscell{\underline{\scorepos{72.6}{+0.2}}} & \poscell{\scorepos{74.1}{+1.5}} \\
& Success-only
& \poscell{\scorepos{81.5}{+1.5}} & \poscell{\scorepos{77.5}{+1.9}} & \poscell{\underline{\scorepos{77.8}{+1.9}}} & \poscell{\scorepos{75.2}{+3.2}}
& \poscell{\scorepos{81.0}{+2.0}} & \poscell{\underline{\scorepos{77.8}{+0.9}}} & \negcell{\scoreneg{75.3}{-1.0}} & \poscell{\scorepos{74.9}{+1.2}}
& \failcell{\scorezero{74.5}{+0.0}} & \failcell{\scoreneg{72.0}{-1.0}} & \failcell{\scorepos{72.5}{+0.1}} & \failcell{\scorepos{73.4}{+0.8}} \\
\midrule
\multirow{4}{*}{\textbf{OfficeQA}} & \mbox{Parent ($r=0$)}
& \scorebase{77.6} & \scorebase{62.8} & \scorebase{59.2} & \scorebase{59.4}
& \zerocell{\selcell{\underline{\scorebase{77.6}}}} & \zerocell{\selcell{\scorebase{65.5}}} & \zerocell{\selcell{\scorebase{68.8}}} & \zerocell{\selcell{\underline{\scorebase{73.1}}}}
& \zerocell{\selcell{\underline{\scorebase{73.5}}}} & \zerocell{\selcell{\underline{\scorebase{56.8}}}} & \zerocell{\selcell{\underline{\scorebase{52.5}}}} & \zerocell{\selcell{\underline{\scorebase{58.6}}}} \\
& Normal
& \poscell{\selcell{\underline{\scorepos{87.8}{+10.2}}}} & \poscell{\selcell{\underline{\scorepos{69.6}{+6.8}}}} & \poscell{\selcell{\underline{\scorepos{70.5}{+11.3}}}} & \poscell{\selcell{\underline{\scorepos{71.1}{+11.7}}}}
& \failcell{\scorezero{77.6}{+0.0}} & \failcell{\scorezero{65.5}{+0.0}} & \failcell{\underline{\scorepos{69.9}{+1.1}}} & \failcell{\scoreneg{71.0}{-2.1}}
& \failcell{\scorezero{73.5}{+0.0}} & \failcell{\scoreneg{54.7}{-2.0}} & \failcell{\scoreneg{48.4}{-4.2}} & \failcell{\scoreneg{52.7}{-5.9}} \\
& Fail-only
& \poscell{\scorepos{85.7}{+8.2}} & \poscell{\underline{\scorepos{69.6}{+6.8}}} & \poscell{\scorepos{67.2}{+8.0}} & \poscell{\scorepos{67.7}{+8.4}}
& \failcell{\scorezero{77.6}{+0.0}} & \failcell{\scoreneg{64.9}{-0.7}} & \failcell{\scorepos{69.1}{+0.2}} & \failcell{\scoreneg{70.8}{-2.2}}
& \failcell{\scorezero{73.5}{+0.0}} & \failcell{\scoreneg{52.7}{-4.1}} & \failcell{\scoreneg{50.4}{-2.2}} & \failcell{\scoreneg{53.7}{-4.9}} \\
& Success-only
& \poscell{\scorepos{85.7}{+8.2}} & \poscell{\scorepos{66.9}{+4.1}} & \poscell{\scorepos{64.1}{+4.9}} & \poscell{\scorepos{63.9}{+4.5}}
& \failcell{\scorezero{77.6}{+0.0}} & \failcell{\underline{\scorepos{67.6}{+2.0}}} & \failcell{\scoreneg{68.1}{-0.8}} & \failcell{\scoreneg{67.9}{-5.1}}
& \failcell{\scorezero{73.5}{+0.0}} & \failcell{\underline{\scorezero{56.8}{+0.0}}} & \failcell{\scoreneg{49.1}{-3.5}} & \failcell{\scoreneg{55.3}{-3.3}} \\
\midrule
\multirow{4}{*}{\textbf{SpreadsheetBench}} & \mbox{Parent ($r=0$)}
& \scorebase{41.0} & \scorebase{50.2} & \scorebase{44.5} & \scorebase{49.8}
& \scorebase{38.5} & \scorebase{42.0} & \scorebase{41.2} & \scorebase{51.3}
& \scorebase{38.5} & \scorebase{39.9} & \scorebase{39.9} & \scorebase{44.4} \\
& Normal
& \poscell{\underline{\scorepos{82.1}{+41.0}}} & \poscell{\scorepos{82.2}{+32.0}} & \poscell{\scorepos{79.8}{+35.3}} & \poscell{\scorepos{77.6}{+27.8}}
& \poscell{\selcell{\underline{\scorepos{82.1}{+43.6}}}} & \poscell{\selcell{\scorepos{79.7}{+37.7}}} & \poscell{\selcell{\underline{\scorepos{85.1}{+43.8}}}} & \poscell{\selcell{\scorepos{76.8}{+25.5}}}
& \poscell{\selcell{\underline{\scorepos{76.9}{+38.5}}}} & \poscell{\selcell{\underline{\scorepos{68.7}{+28.8}}}} & \poscell{\selcell{\underline{\scorepos{72.2}{+32.3}}}} & \poscell{\selcell{\underline{\scorepos{65.5}{+21.1}}}} \\
& Fail-only
& \poscell{\selcell{\underline{\scorepos{82.1}{+41.0}}}} & \poscell{\selcell{\underline{\scorepos{85.8}{+35.6}}}} & \poscell{\selcell{\underline{\scorepos{80.0}{+35.5}}}} & \poscell{\selcell{\underline{\scorepos{77.9}{+28.1}}}}
& \poscell{\scorepos{76.9}{+38.5}} & \poscell{\underline{\scorepos{81.5}{+39.5}}} & \poscell{\scorepos{83.7}{+42.5}} & \poscell{\underline{\scorepos{78.5}{+27.1}}}
& \poscell{\scorepos{69.2}{+30.8}} & \poscell{\scorepos{64.8}{+24.9}} & \poscell{\scorepos{66.4}{+26.5}} & \poscell{\underline{\scorepos{65.5}{+21.1}}} \\
& Success-only
& \failcell{\scorezero{41.0}{+0.0}} & \failcell{\scoreneg{47.7}{-2.5}} & \failcell{\scoreneg{41.6}{-2.9}} & \failcell{\scorepos{49.9}{+0.1}}
& \poscell{\scorepos{48.7}{+10.3}} & \poscell{\scorepos{50.2}{+8.2}} & \poscell{\scorepos{48.4}{+7.1}} & \poscell{\scorepos{54.7}{+3.3}}
& \poscell{\scorepos{48.7}{+10.3}} & \negcell{\scoreneg{39.1}{-0.7}} & \poscell{\scorepos{41.0}{+1.1}} & \poscell{\scorepos{46.0}{+1.6}} \\
\midrule
\multirow{4}{*}{\textbf{LiveMath}} & \mbox{Parent ($r=0$)}
& \scorebase{51.4} & \underline{\scorebase{49.1}} & \scorebase{49.6} & \underline{\scorebase{62.5}}
& \scorebase{42.9} & \scorebase{41.5} & \scorebase{47.8} & \scorebase{62.5}
& \scorebase{40.0} & \scorebase{10.4} & \scorebase{12.4} & \scorebase{16.7} \\
& Normal
& \failcell{\scorezero{51.4}{+0.0}} & \failcell{\scoreneg{40.6}{-8.5}} & \failcell{\scoreneg{47.9}{-1.7}} & \failcell{\scoreneg{54.2}{-8.3}}
& \poscell{\selcell{\underline{\scorepos{74.3}{+31.4}}}} & \poscell{\selcell{\underline{\scorepos{64.2}{+22.6}}}} & \poscell{\selcell{\underline{\scorepos{66.4}{+18.5}}}} & \poscell{\selcell{\underline{\scorepos{79.2}{+16.7}}}}
& \selcell{\underline{\scorepos{54.3}{+14.3}}} & \selcell{\scorepos{20.8}{+10.4}} & \selcell{\underline{\scorepos{20.6}{+8.2}}} & \selcell{\underline{\scorepos{50.0}{+33.3}}} \\
& Fail-only
& \poscell{\selcell{\underline{\scorepos{57.1}{+5.7}}}} & \negcell{\selcell{\scoreneg{42.5}{-6.6}}} & \poscell{\selcell{\underline{\scorepos{57.8}{+8.3}}}} & \negcell{\selcell{\scoreneg{45.8}{-16.7}}}
& \poscell{\underline{\scorepos{74.3}{+31.4}}} & \poscell{\scorepos{56.6}{+15.1}} & \poscell{\scorepos{63.0}{+15.1}} & \zerocell{\scorezero{62.5}{+0.0}}
& \poscell{\scorepos{48.6}{+8.6}} & \poscell{\underline{\scorepos{28.3}{+17.9}}} & \poscell{\scorepos{20.0}{+7.6}} & \poscell{\scorepos{41.7}{+25.0}} \\
& Success-only
& \failcell{\scorezero{51.4}{+0.0}} & \failcell{\scoreneg{43.4}{-5.7}} & \failcell{\scoreneg{48.1}{-1.4}} & \failcell{\scoreneg{45.8}{-16.7}}
& \poscell{\scorepos{62.9}{+20.0}} & \poscell{\scorepos{61.3}{+19.8}} & \poscell{\scorepos{61.9}{+14.0}} & \poscell{\scorepos{66.7}{+4.2}}
& \failcell{\scorezero{40.0}{+0.0}} & \failcell{\scorepos{23.6}{+13.2}} & \failcell{\scorepos{14.4}{+2.0}} & \failcell{\scorepos{25.0}{+8.3}} \\
\midrule
\multirow{4}{*}{\textbf{DocVQA}} & \mbox{Parent ($r=0$)}
& \zerocell{\selcell{\underline{\scorebase{96.2}}}} & \zerocell{\selcell{\underline{\scorebase{92.0}}}} & \zerocell{\selcell{\underline{\scorebase{87.9}}}} & \zerocell{\selcell{\underline{\scorebase{93.7}}}}
& \scorebase{92.5} & \scorebase{95.5} & \scorebase{94.8} & \scorebase{96.9}
& \notavailablecell & \notavailablecell & \notavailablecell & \notavailablecell \\
& Normal
& \failcell{\scorezero{96.2}{+0.0}} & \failcell{\scoreneg{91.7}{-0.3}} & \failcell{\scoreneg{75.4}{-12.4}} & \failcell{\scoreneg{91.2}{-2.5}}
& \poscell{\selcell{\underline{\scorepos{94.3}{+1.9}}}} & \poscell{\selcell{\underline{\scorepos{96.0}{+0.5}}}} & \poscell{\selcell{\underline{\scorepos{95.1}{+0.3}}}} & \negcell{\selcell{\scoreneg{96.6}{-0.3}}}
& \notavailablecell & \notavailablecell & \notavailablecell & \notavailablecell \\
& Fail-only
& \failcell{\scorezero{96.2}{+0.0}} & \failcell{\scoreneg{91.7}{-0.3}} & \failcell{\scoreneg{76.8}{-11.1}} & \failcell{\scoreneg{89.5}{-4.3}}
& \poscell{\underline{\scorepos{94.3}{+1.9}}} & \negcell{\scoreneg{95.2}{-0.3}} & \poscell{\scorepos{95.0}{+0.2}} & \poscell{\underline{\scorepos{97.0}{+0.1}}}
& \notavailablecell & \notavailablecell & \notavailablecell & \notavailablecell \\
& Success-only
& \failcell{\scorezero{96.2}{+0.0}} & \failcell{\scoreneg{91.2}{-0.8}} & \failcell{\scoreneg{75.3}{-12.6}} & \failcell{\scoreneg{91.0}{-2.7}}
& \failcell{\scorezero{92.5}{+0.0}} & \failcell{\scorepos{95.7}{+0.3}} & \failcell{\scorezero{94.8}{+0.0}} & \failcell{\scoreneg{96.8}{-0.1}}
& \notavailablecell & \notavailablecell & \notavailablecell & \notavailablecell \\
\bottomrule
\end{tabularx}
}
\caption{Feedback-conditioned results across models and benchmarks. Cells report score (change from the corresponding model--benchmark parent, pp). Blue shading and boldface mark validation-selected views; teal, orange, and gray indicate positive, negative, and zero, parent-identical, or unavailable comparisons; underlining marks within-setting metric bests. DeepSeek--DocVQA is unsupported. Scores use benchmark-specific evaluators; R/T are equal-weight averages over available probes.}
\label{tab:main-results}
\end{table*}

\subsection{Experimental Setup}

\paragraph{Benchmarks and evolution protocol.}
We evaluate three underlying models across five benchmarks. GPT-5.5 and Gemini 3.1 Pro are evaluated on SearchQA, OfficeQA, SpreadsheetBench, LiveMath, and DocVQA. We exclude DeepSeek--DocVQA because its endpoint does not accept the benchmark's native page images, and adding an OCR stage would alter the task interface. This yields 14 model--benchmark settings and 42 matched feedback runs. ALFWorld is reported in Appendix~\ref{tab:detailed-scores}.

For each model--benchmark setting, Normal, Fail-only, and Success-only start from the same parent skill and use the same executor and optimizer configuration, revision procedure, validation rule, and ten-round budget. They differ only in the trajectories shown to the optimizer: both successful and failed trajectories (Normal), failed trajectories only, or successful trajectories only. Each round executes 40 training trajectories (36 for LiveMath); a nonempty feedback view triggers one optimizer call proposing at most four minimal, task-general edits, followed by full-split validation. A candidate becomes the next-round skill when validation does not decrease; otherwise, the current skill is retained. Only a strict validation improvement updates the best checkpoint. Runs stop after ten rounds, five consecutive regressions or no-ops, or an empty feedback pool; across rounds, only the incumbent skill carries forward. Appendices~\ref{tab:benchmark-protocol}, \ref{tab:evolution-protocol}, \ref{tab:execution-identity-controls}, and~\ref{tab:evolution-run-summary} provide the benchmark definitions, common protocol, execution controls, and per-run records.

\begin{figure*}[!t]
\centering
\includegraphics[width=\textwidth]{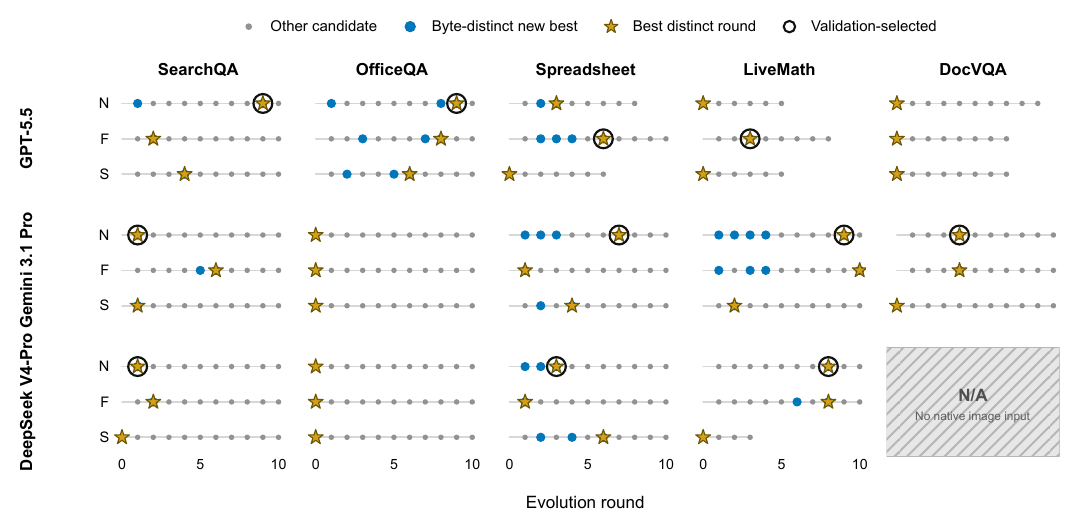}
\caption{Evolution events across 14 model--benchmark settings. Gray dots mark other evaluated candidates, blue dots mark byte-distinct validation bests, stars mark the best byte-distinct round within each feedback view, and black rings mark benchmark-level selections among the three views. Rows denote models, columns denote benchmarks, and N, F, and S denote Normal, Fail-only, and Success-only. DeepSeek--DocVQA is unavailable because its endpoint does not accept native page images.}
\label{fig:lifecycle-attribution}
\end{figure*}

\paragraph{Robustness and transfer probes.}
We assess same-task robustness with three tiers: R1 changes surface form or input representation while preserving the task and gold answer, R2 adds irrelevant context or distractors, and R3 changes the output or interface contract. Transfer also has three tiers: T1 uses a new official task from the same source artifact, T2 uses a new source within a task subtype represented in the original evaluation panel, and T3 uses a new source with a subtype, difficulty, or distribution shift. Each probe score averages three deployments on a fixed panel. R and T are equal-weight macro-averages over the available tiers. Appendix~\ref{tab:probe-instantiations} lists the benchmark-specific probes and eligible sample sizes.

\paragraph{Test-time-scaling controls.}
Prior work studies parallel and sequential test-time scaling for agents \cite{zhu2025scalingtesttimeagents,kim2026scalingagenticcoding,li2026benchmarktesttimescaling}, including comparisons with harness evolution \cite{wang2026rethinking}. Across five GPT-5.5 benchmarks, we compare one-call deployment of the validation-selected skill with two controls initialized from the frozen parent skill for each task. At budget $K$, which includes the parent attempt, Parallel Sampling reports oracle any-success over that attempt and $K-1$ independent attempts, while Sequential Refinement conditions each additional attempt only on the task and preceding response and reports the final response. The controls share the executor, tool interface, test pool, verifier, and call budget; curves compare items receiving the full budget. SearchQA and SpreadsheetBench are the primary contrasts. Appendices~\ref{tab:tts-differences}, \ref{tab:tts-definition-budget}, \ref{tab:scaling-numeric}, and~\ref{fig:searchqa-tts-amortization} provide complete definitions, allocations, scores, and the call accounting for the selected SearchQA run.

\subsection{Main Results}

\noindent\textbf{All selected evolved skills come from feedback containing failed trajectories; Success-only is never selected.}
Across the 14 model--benchmark settings in Table~\ref{tab:main-results}, validation selects 11 evolved skills. Normal accounts for nine selections and Fail-only for two, while the parent is retained in three settings: Gemini--OfficeQA, DeepSeek--OfficeQA, and GPT-5.5--DocVQA. On the released test, nine of the 11 selected evolved skills improve over their parents, with positive gains ranging from $0.5$ to $37.7$ points. SpreadsheetBench shows the largest and most consistent gains: $35.6$ points for GPT-5.5, $37.7$ for Gemini, and $28.8$ for DeepSeek. LiveMath is more variable: GPT-5.5 declines by $6.6$ points, whereas Gemini and DeepSeek improve by $22.6$ and $10.4$ points, respectively. SearchQA changes are smaller: $+2.3$, $-0.4$, and $+0.9$ points for GPT-5.5, Gemini, and DeepSeek, respectively. OfficeQA selects an evolved skill only for GPT-5.5 ($+6.8$ points), with Gemini and DeepSeek retaining their parents; for DocVQA, Gemini selects Normal ($+0.5$), GPT-5.5 retains its parent, and DeepSeek is unavailable.

\noindent\textbf{Accepted revisions are sparse, and selected rounds vary across models, benchmarks, and feedback views.}
Figure~\ref{fig:lifecycle-attribution} summarizes 388 evaluated candidates across 42 observed feedback trajectories, of which 55 establish byte-distinct validation bests. SearchQA selects an evolved skill for all three models, at round 9 for GPT-5.5 and round 1 for both Gemini and DeepSeek. SpreadsheetBench also selects an evolved skill for all three models, at rounds 6, 7, and 3, respectively. OfficeQA selects a round-9 GPT-5.5 skill but retains the parent for Gemini and DeepSeek. DocVQA retains the GPT-5.5 parent and selects a round-4 Gemini skill, while DeepSeek is unavailable. LiveMath selects round-3, round-9, and round-8 skills for GPT-5.5, Gemini, and DeepSeek, respectively. Across all 14 model--benchmark settings, 11 select an evolved skill and three retain the parent; five evolved selections occur in rounds 1--4 and six in rounds 6--9.

\noindent\textbf{Across the 11 settings with a validation-selected evolved skill, robustness and transfer each improve in nine, and seven improve both.}
Complete test, robustness, and transfer results are available for all 14 model--benchmark settings. Seven of the 11 evolved selections improve both robustness and transfer. SpreadsheetBench is positive on test, robustness, and transfer for all three models. SearchQA improves transfer for all three models, while robustness changes by $+0.2$, $-0.3$, and $-0.2$ points for GPT-5.5, Gemini, and DeepSeek. LiveMath improves all three measures for Gemini and DeepSeek; GPT-5.5 instead gains $8.3$ points in robustness while losing $6.6$ on test and $16.7$ on transfer. Gemini--DocVQA gains $0.5$ points on test and $0.3$ on robustness but loses $0.3$ on transfer. GPT-5.5--OfficeQA improves all three measures by $6.8$, $11.3$, and $11.7$ points.

\FloatBarrier
\section{Analysis}
\label{sec:analysis}

\begin{samepage}
\subsection{Cross-Model Evidence on SearchQA}

SearchQA provides the broadest cross-model coverage. Appendix~\ref{tab:cross-model-results} extends the comparison beyond GPT-5.5 to seven models with complete released-test results: Claude Opus, Kimi, Gemini, DeepSeek, GLM, Grok, and Qwen3.5-Plus. Across these models, 191 of 210 candidates change the incoming skill, but only 29 establish byte-distinct validation bests. This gap separates revision activity from retained progress: modification is common, whereas validation-confirmed improvement is sparse. Selected rounds span 1--7, compared with round 9 for GPT-5.5, so selection timing is not uniform across models. Six of the seven validation-selected skills improve released-test performance by 0.86--15.07 points, while Gemini decreases by $0.36$ points. Test-best and validation-selected views differ in six settings, indicating model-dependent ranking shifts between validation and test. Qwen3.5-Plus also improves robustness by $2.19$ points and transfer by $2.55$ points. Across all eight SearchQA models, seven validation-selected skills improve released-test performance, while the selected view, round, and validation-to-test ranking differ by model.
\end{samepage}

\subsection{Generalization of Evolved Skills}
\label{sec:behavioral-scope}

Cross-model evidence establishes breadth across model families. The robustness and transfer results in Table~\ref{tab:main-results} and Appendix~\ref{tab:selected-scope-crossmodel} test whether the retained behavior extends beyond the released-test distribution. They capture different consequences of a selected skill rather than interchangeable forms of improvement. SpreadsheetBench improves released-test, robustness, and transfer performance for all three models, whereas SearchQA improves transfer in all three models with complete diagnostics but leaves robustness at or below the parent. Gemini--DocVQA shows small released-test and robustness gains but a small transfer decrease. LiveMath is model dependent: GPT-5.5 improves robustness but loses released-test and transfer performance, whereas DeepSeek and Gemini improve all three; Gemini gains $22.6$ points on released test, $18.5$ on robustness, and $16.7$ on transfer. Generalization therefore depends on the interaction among the benchmark, model, and retained skill, not on validation improvement alone.

The GPT-5.5 probe-level results in Appendix~\ref{tab:task-scope-exact} further distinguish local robustness from transfer. SpreadsheetBench remains positive across R1--R3 and T2, with the transfer gain narrowing to $+2.9$ on T3. LiveMath also improves across all robustness probes, but its transfer differences are $0.0$ on T2 and $-33.3$ on T3. The four- and two-item transfer panels identify probe-level effects, while the 106-item released test provides broader evidence for the GPT-5.5--LiveMath validation-to-test reversal. Together, these probes characterize benchmark-specific generalization across the evaluated robustness and transfer conditions.

\subsection{Effects of Feedback Composition}

Tables~\ref{tab:main-results} and~\ref{tab:feedback-improvement-counts} show that no feedback view has a fixed advantage across model--benchmark settings. Fail-only produces a byte-distinct validation improvement in 11 of 14 settings, compared with 10 for Normal and six for Success-only. Normal is nevertheless selected most often: nine settings, versus two for Fail-only and none for Success-only. Normal and Fail-only each improve released-test performance in nine settings. Fail-only improves robustness and transfer in nine of 14 settings, compared with eight and nine for Normal. Success-only improves robustness and transfer in five and six settings, respectively, and improves fewer settings on every reported metric. These view-specific counts exclude parent-identical branches, and a setting may contribute to more than one row.

\begin{table}[t]
\centering
\scriptsize
\setlength{\tabcolsep}{2.5pt}
\renewcommand{\arraystretch}{1.08}
\begin{tabular*}{\columnwidth}{@{\extracolsep{\fill}}lccccc@{}}
\toprule
\textbf{View} & \textbf{Selected} & \textbf{Val.$\uparrow$} &
\textbf{Test$\uparrow$} & \textbf{Robust.$\uparrow$} & \textbf{Transfer$\uparrow$} \\
\midrule
Normal & 9 & 10 & 9 & 8 & 9 \\
Fail-only & 2 & 11 & 9 & 9 & 9 \\
Success-only & 0 & 6 & 5 & 5 & 6 \\
\bottomrule
\end{tabular*}
\caption{Counts of model--benchmark settings improved or selected by each feedback view.}
\label{tab:feedback-improvement-counts}
\end{table}

The artifact audits in Appendices~\ref{tab:selected-skill-cards}, \ref{tab:officeqa-feedback-case}, and~\ref{tab:spreadsheet-feedback-case} clarify the contrast between Normal and Fail-only. On OfficeQA, failures expose unsupported first-turn answers, while successful traces give Normal a reference for what should be preserved. The resulting skill combines a retrieve--read--compute rule with answer-format constraints and exceeds Fail-only on validation, robustness, and transfer. On SpreadsheetBench, Fail-only focuses on a verifier-visible defect: formula strings leave required cells empty. Its explicit write--reopen--check procedure yields the strongest test, robustness, and transfer results. Normal can use successes to broaden a repair, whereas Fail-only targets the defect.

Appendix~\ref{tab:success-only-cases} explains Success-only's rarity. \mbox{Successes show} what worked but provide no direct contrast for identifying what must be corrected. When only a few successes are available, the optimizer can mistake incidental commonalities for task-level rules. In DeepSeek--LiveMath, the round-3 candidate extrapolates from eight successful traces, adds a ``strongest/equivalence'' heuristic, and changes the required output from an option label to the full option text. Validation falls from 40.0 to 11.4, and the arm ultimately retains the parent. Opus and Qwen3.5-Plus are the two Success-only selections among the eight SearchQA models. Opus's successful traces support a repeated task-wide specification---identify the clue referent and return a short canonical answer---and the round-7 skill improves validation from 77.0 to 79.0 and released-test performance by 3.79 points. Qwen3.5-Plus selects its round-3 skill, improving validation from 73.0 to 76.5 and released-test performance by 2.93 points. Together, these cases suggest that Success-only is most useful when positive trajectories expose a stable shared specification, whereas negative contrast helps distinguish task-level rules from incidental patterns.

\subsection{Effects of Additional Evolution Rounds}
\label{sec:evolution-pathway}

Across the 42 primary runs summarized in Appendix~\ref{fig:lifecycle-heatmap}, additional rounds expand the search horizon, but their average yield falls after the early stage. Thirty-eight of the 55 byte-distinct validation bests occur in rounds 1--4. The remaining 17 arise from 221 candidates evaluated after round 4. Late search remains consequential: six of the 11 selected evolved skills first appear in rounds 6--9, compared with five in rounds 1--4. A four-round budget would capture most new-best events but miss most final evolved selections.

The late gains do not arise from steady improvement. SearchQA Normal first reaches 81.0 at round 1, then evaluates seven candidates without exceeding it before reaching 82.0 at round 9. OfficeQA Normal improves at rounds 1, 8, and 9, with regressions between them. SpreadsheetBench shows saturation: Normal peaks at round 3 and all five later candidates score lower, while Fail-only reaches its final best at round 6 and then produces four lower-scoring candidates. DocVQA never exceeds its 96.2 parent across 23 candidates.

These trajectories characterize evolution as validation-filtered search rather than monotonic refinement. Additional rounds can uncover a late skill after many rejected candidates, but after saturation they add search cost without retained improvement. Rollback preserves the best validation result; it does not create progress or guarantee test improvement. LiveMath makes this distinction explicit: its round-3 Fail-only skill is the only validation improvement, yet it reduces released-test performance by 6.6 points. Round budgets should therefore be evaluated by the timing and frequency of new bests, not by the number of revisions alone. These diminishing late-round returns motivate a complementary question: whether additional per-instance inference can recover the gains produced by persistent skill evolution.

\begin{figure*}[!t]
\centering
\includegraphics[width=\textwidth]{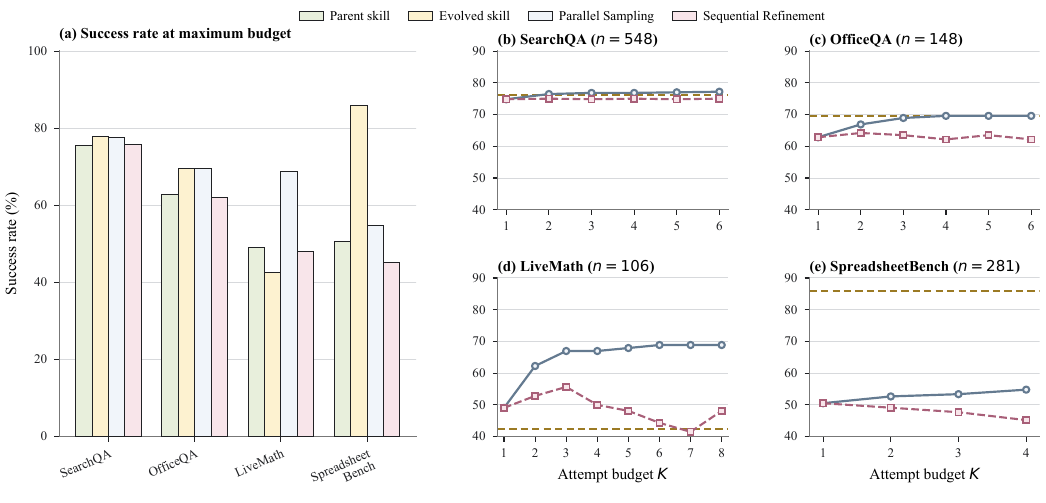}
\caption{GPT-5.5 test-time scaling on benchmarks with byte-distinct evolved skills. (a) One-call parent and evolved-skill performance compared with maximum-budget Parallel Sampling and Sequential Refinement. (b--e) Common-support budget curves; horizontal references show one-call evolved-skill performance. Parallel reports oracle any-success, and Sequential reports the final attempt.}
\label{fig:tts-composite}
\end{figure*}

\subsection{Self-Evolution versus Test-Time Scaling}
\label{sec:test-time-scaling}

Figure~\ref{fig:tts-composite} shows that oracle Parallel Sampling nearly matches the evolved skill on SearchQA but recovers little of its advantage on SpreadsheetBench. On SearchQA, the frozen parent execution scores 75.64, the evolved skill 77.93, Parallel 77.50, and Sequential 75.79; Parallel is therefore only 0.43 points below the evolved skill. On SpreadsheetBench, the corresponding scores are 50.53, 85.77, 54.80, and 45.20, leaving a 30.96-point gap between evolution and Parallel.

This contrast separates response diversity from persistent policy change. The SearchQA skill mainly adds answer-form guidance---answer type, canonical naming, and concise output---so independent samples can explore alternatives and oracle selection recovers most of the modest gain. SpreadsheetBench instead requires a multi-step workflow: inspect the workbook, execute a self-contained script, materialize values, save the output, and verify target cells. The 30.96-point gap indicates that additional parent-skill samples rarely reproduce this complete procedure. Sequential Refinement also remains near the parent on SearchQA and OfficeQA and declines on SpreadsheetBench, showing that conditioning on a preceding response is not itself corrective feedback.

\enlargethispage{\baselineskip}
OfficeQA provides a second sampling-recoverable case: maximum-budget Parallel matches the evolved score at 69.59. On LiveMath, the validation-selected skill scores 42.45 on test, the parent 49.06, and Parallel 68.87; this comparison combines a sampling gain with a validation-to-test reversal. Across these controls, test-time scaling can exploit response diversity available under the parent skill, while persistent skill improvement reflects retained changes across task instances. Oracle score parity and persistent skill improvement therefore measure different outcomes.

\FloatBarrier
\section{Conclusion}

Persistent skill self-evolution behaves as sparse search rather than steady improvement. Across 42 feedback runs in 14 model--benchmark settings, only 55 of 388 candidates establish byte-distinct validation bests, and validation selects an evolved skill in 11 settings. Selected skills appear in early and late rounds, while other runs saturate or retain the parent. Normal accounts for nine selected skills and Fail-only for two, so all 11 evolved selections use failed trajectories; Success-only is never selected in the primary study, while the leading failure-containing view varies across models and benchmarks. A broader SearchQA analysis across eight models shows similarly sparse, feedback-dependent dynamics. Selected skills improve released-test performance, robustness, and transfer in nine settings each, with seven improving both robustness and transfer. In the GPT-5.5 test-time-scaling controls, oracle Parallel Sampling comes within 0.43 points of the evolved SearchQA skill but remains 30.96 points behind on SpreadsheetBench, while Sequential Refinement reproduces neither gain. These findings characterize self-evolution as feedback-conditioned, validation-filtered search over persistent skills and show why endpoint scores alone are insufficient. Evaluations should report search trajectories, skill identity, downstream generalization, and explicit test-time-compute controls.

\section{Limitations}

Although the primary study covers 14 model--benchmark settings across five heterogeneous benchmarks, its coverage of agent-skill settings remains incomplete. In particular, we do not evaluate on dedicated skill benchmarks such as SkillsBench and SkillLearnBench \cite{li2026skillsbench,zhong2026skilllearnbench}, which cover broader skill-dependent tasks and continual skill generation. Future work can apply our matched feedback-isolation and validation-selection protocol to these benchmarks to test whether the observed feedback rankings and sparse search dynamics generalize beyond the present task suite.

\FloatBarrier
\raggedbottom
\bibliography{references}

\appendix
\onecolumn
\raggedbottom
\setlength{\floatsep}{14pt plus 2pt minus 2pt}
\setlength{\textfloatsep}{14pt plus 2pt minus 2pt}
\setlength{\intextsep}{12pt plus 2pt minus 2pt}
\setlength{\abovecaptionskip}{6pt}
\setlength{\belowcaptionskip}{0pt}
\renewcommand{\thetable}{A\arabic{table}}
\setcounter{table}{0}
\renewcommand{\thefigure}{A\arabic{figure}}
\setcounter{figure}{0}

\section{Experimental Protocol and Evaluation Design}
\label{app:additional-experimental-details}
\subsection{Benchmarks, Metrics, and Data Splits}

\begin{table}[H]
\centering
\footnotesize
\setlength{\tabcolsep}{2.2pt}
\renewcommand{\arraystretch}{1.08}
\begin{tabularx}{0.99\textwidth}{@{}L{0.15\textwidth} L{0.17\textwidth} L{0.20\textwidth} L{0.27\textwidth} Y@{}}
\toprule
\textbf{Benchmark} & \textbf{Task and output} & \textbf{Source and fixed pool} & \textbf{Primary metric} & \textbf{Auxiliary metric} \\
\midrule
SearchQA & Short-answer QA over retrieved web snippets & SearchQA \cite{dunn2017searchqa}; fixed SkillOpt-Lite subset (400/200/1,400) & 1 iff the normalized prediction exactly matches any reference & Maximum token F1 \\
OfficeQA & Grounded QA over U.S.\ Treasury Bulletins with search/read tools & OfficeQA Full; complete 246-item release (49/49/148) & 1 iff the adapter-normalized prediction exactly matches the target & Token F1 \\
SpreadsheetBench & Instruction-following spreadsheet manipulation & Verified-400 \cite{ma2024spreadsheetbench} (80/39/281) & 1 iff every required workbook test case passes & Fraction of test cases passed \\
ALFWorld & Multi-step household task completion in a text environment & ALFWorld \cite{shridhar2021alfworld}; fixed episodes (200/140/134) & 1 iff the simulator returns \texttt{won} & Same binary outcome \\
LiveMath & Multiple-choice reasoning over recent mathematics papers & LiveMathematicianBench \cite{he2026livemathematicianbench}; Nov.\ 2025--Feb.\ 2026 snapshot (36/35/106) & 1 iff the parsed option equals the gold option & Same binary outcome \\
DocVQA & Question answering over document-page images & DocVQA \cite{mathew2021docvqa}; sample from the official validation split (107/53/374) & 1 iff the maximum ANLS across references is at least 0.999 & Maximum ANLS \\
\bottomrule
\end{tabularx}
\normalsize
\caption{Benchmark tasks, sources, fixed evaluation pools, and metrics. ALFWorld is reported as an additional analysis, separately from the primary aggregates.}
\label{tab:benchmark-protocol}
\vspace{0.35em}
\begin{minipage}{0.99\textwidth}
\footnotesize
\textit{Note.} Every reported Score is 100 times the mean of the binary outcome in the fourth column. Auxiliary metrics enter the validation gate or tie-breaking rule when configured, while reported hard scores use the primary metric. OfficeQA uses normalized exact match in our fixed adapter; its official release (\url{github.com/databricks/officeqa}) also provides tolerance-based numerical scoring. The DocVQA hard score thresholds maximum ANLS at 0.999.
\end{minipage}
\end{table}

\subsection{Evolution Protocol and Feedback Isolation}

\begin{table}[H]
\centering
\small
\setlength{\tabcolsep}{3.5pt}
\renewcommand{\arraystretch}{1.00}
\begin{tabularx}{0.99\textwidth}{@{}L{0.19\textwidth} X@{}}
\toprule
\textbf{Protocol component} & \textbf{Fixed setting} \\
\midrule
Run structure &
Each benchmark--view run starts from the same benchmark-specific parent skill and continues for at most ten rounds. \\

Per-round budget &
Each round executes 40 training trajectories, except LiveMath, which uses its full 36-item training split; a nonempty view triggers one optimizer call, up to four edits, and full-split validation. \\

Feedback exposure &
Normal exposes successful and failed trajectories, Fail-only exposes failed trajectories, and Success-only exposes successful trajectories. \\

Revision discipline &
The same revision instruction applies to all three feedback views: make at most four minimal, task-general edits supported by visible evidence. \\

Acceptance and stopping &
A candidate becomes the next-round skill when validation does not decrease; otherwise, the incumbent is retained. Only a strict validation improvement updates the best checkpoint. A run stops after ten rounds, five consecutive regressions or no-ops, or an empty view-specific feedback pool; no optimizer proposal is made in the last case. \\

Selection and identity &
The validation-best skill is restored for each run. Cross-view reporting uses validation hard score and validation soft score; candidate and retained artifacts are compared byte-for-byte. \\
\bottomrule
\end{tabularx}
\normalsize
\caption{Common multi-round evolution protocol.}
\label{tab:evolution-protocol}
\end{table}

\begin{table}[H]
\centering
\footnotesize
\setlength{\tabcolsep}{3.0pt}
\renewcommand{\arraystretch}{1.06}
\begin{tabularx}{\textwidth}{@{}L{0.17\textwidth} L{0.25\textwidth} Y@{}}
\toprule
\textbf{Control} & \textbf{Setting} & \textbf{Matched scope} \\
\midrule
Underlying model &
GPT-5.5; \texttt{gemini-3.1-pro-preview}; \texttt{deepseek-v4-pro} &
Within each setting, executor and optimizer use the same model. Across rounds, only the incumbent skill carries over; earlier trajectories and rejected candidates do not. \\

Execution pathway &
Fixed within each model--benchmark setting &
All three feedback views use the same executor and tool interface within a matched setting. \\

Scheduling &
GPT-5.5: 20 workers on SearchQA, 8 on SpreadsheetBench, and 4 on the other three benchmarks; Gemini and DeepSeek: 50 Stage-1 lanes &
Parallelism affects scheduling only; per-round batch sizes, the optimizer-call cap, and the validation gate are defined in Table~\ref{tab:evolution-protocol}. \\

Randomness and retry &
Fixed splits, item identities, and commands; infrastructure-only completion retries &
Seeds determine loader or panel order where exposed, while API generation may vary. Behavior failures remain scored outcomes, and infrastructure-incomplete items are retried under the same protocol. \\

Matched optimizer contract &
Same parent, optimizer instructions, sample schema, task panel, edit cap, and validation gate &
Within each setting, the optimizer prompt is fixed across rounds and differs across views only in its visible-directory declaration. Each call receives the incumbent skill and the permitted training records. Every record contains the task input, the executor's final output, and its success label and score. Records also include the execution trace when available; failed records retain verifier diagnostics, including expected--observed mismatches when available. \\

Arm-specific evidence &
Normal: passed and failed records; Fail-only: failed records; Success-only: passed records &
All eligible records from each configured round batch are exposed without balancing or resampling; only visibility changes, and neither records nor evolving state cross views or settings. \\

Selection and evaluation boundary &
Validation hard score, then the configured auxiliary score; SHA-256 artifact identity &
Validation is external to the optimizer and drives acceptance and cross-view selection. Post-selection evaluation comprises test, R/T, TTS, and verifier sensitivity. The optimizer modifies the skill artifact; a new best requires both a strict validation improvement and a distinct skill hash. Byte-identical artifacts share the same skill identity. \\
\bottomrule
\end{tabularx}
\normalsize
\caption{Execution, feedback-isolation, and artifact-identity controls for matched comparisons across the three feedback views.}
\label{tab:execution-identity-controls}
\end{table}

\subsection{Post-Selection Evaluation Design}

\begin{table}[H]
\centering
\small
\setlength{\tabcolsep}{3.2pt}
\renewcommand{\arraystretch}{1.06}
\begin{tabularx}{0.99\textwidth}{L{0.17\textwidth} C L{0.30\textwidth} L{0.28\textwidth}}
\toprule
\textbf{Benchmark} & \textbf{Train/Val/Test} & \textbf{Post-selection diagnostics} & \textbf{TTS budget} \\
\midrule
SearchQA & 400/200/1,400 & Three-repeat evaluation; R/T; verifier & $n=1{,}400$, $K\leq6$; $+4{,}924$ calls/control \\
OfficeQA & 49/49/148 & Three-repeat evaluation; R/T; verifier & $n=148$, $K\leq6$; $+740$ calls/control \\
ALFWorld & 200/140/134 & R/T & -- \\
SpreadsheetBench & 80/39/281 & Three-repeat evaluation; R/T; verifier & $n=281$, $K\leq4$; $+843$ calls/control \\
LiveMath & 36/35/106 & Three-repeat evaluation; R/T; verifier & $n=106$, $K\leq8$; $+742$ calls/control \\
DocVQA & 107/53/374 & Three-repeat evaluation; R/T; verifier & $n=374$, $K\leq4$; $+1{,}122$ calls/control \\
\bottomrule
\end{tabularx}
\normalsize
\caption{Benchmark splits and GPT-5.5 post-selection evaluation budgets. Diagnostics summarize the completed GPT-5.5 analyses; TTS budget gives the maximum additional target-model calls per control.}
\label{tab:postfreeze-scope}
\end{table}

\begin{table}[H]
\centering
\small
\setlength{\tabcolsep}{3.2pt}
\renewcommand{\arraystretch}{1.08}
\begin{tabularx}{\textwidth}{@{}L{0.10\textwidth} X@{}}
\toprule
\textbf{Tier} & \textbf{Definition} \\
\midrule
R1 & Same task and gold; surface form or input representation changes. \\
R2 & Same task and gold; irrelevant context or distractors are added. \\
R3 & Same task and gold; the output or interface contract changes. \\
T1 & New official task and gold from the same source artifact. \\
T2 & New official task and gold from a new source in a task subtype represented in the base panel. \\
T3 & New official task and gold from a new source with a deliberate subtype, difficulty, or distribution shift. \\
\bottomrule
\end{tabularx}
\normalsize
\caption{Definitions of robustness and transfer tiers; unavailable tiers are omitted from macro-averages.}
\label{tab:probe-definitions}
\end{table}

\begin{table}[H]
\centering
\footnotesize
\setlength{\tabcolsep}{2.6pt}
\renewcommand{\arraystretch}{1.08}
\begin{tabularx}{\textwidth}{@{}L{0.15\textwidth} L{0.40\textwidth} Y@{}}
\toprule
\textbf{Benchmark} & \textbf{Same-task robustness probes} & \textbf{Official-task transfer probes} \\
\midrule
SearchQA & R1: semantic question paraphrase ($n=97$); R2: answer-excluding irrelevant documents with context-order perturbation ($n=100$); R3: strict JSON answer contract ($n=100$). & T1: unavailable in the fixed pool; T2: new-source QA subtype represented in the base panel ($n=100$); T3: new-source QA subtype absent from the base panel ($n=9$). \\
OfficeQA & R1: instruction reframing with the same question and source ($n=57$); R2: answer-excluding irrelevant documents ($n=50$); R3: strict single-key JSON answer ($n=57$). & T1: new official question on the same document ($n=5$); T2: new-source easy-difficulty items ($n=33$); T3: new-source hard-difficulty items ($n=47$). \\
SpreadsheetBench & R1: task-ID and workbook-path renaming ($n=99$); R2: irrelevant request from another workbook ($n=100$); R3: fixed \texttt{transform(input\_path, output\_path)} response contract ($n=100$). & T1: unavailable in the fixed pool; T2: new-workbook cell-level manipulation ($n=100$); T3: new-workbook sheet-level manipulation ($n=57$). \\
LiveMath & R1: symbol alpha-renaming ($n=97$) and choice-order/label permutation ($n=100$); R2: irrelevant excerpt from another paper ($n=100$); R3: strict single-key JSON answer ($n=100$). & T1: unavailable in the fixed pool; T2: new-paper theorem-type combination represented in the base panel ($n=4$); T3: new-paper theorem-type combination absent from the base panel ($n=2$). \\
DocVQA & R1: instruction reframing with the same question and page ($n=100$); R2: irrelevant page paired with the target page ($n=100$); R3: strict single-key JSON answer ($n=100$). & T1: new official question on the same page ($n=18$); T2: new-page dominant \emph{what}-question subtype ($n=100$); T3: new-page non-dominant interrogative subtype ($n=100$). \\
\bottomrule
\end{tabularx}
\normalsize
\caption{Benchmark-specific robustness and transfer probes. Reported $n$ values are eligible items per condition and deployment.}
\label{tab:probe-instantiations}
\end{table}

\section{Complete Evolution and Endpoint Results}
\label{app:complete-results}
\subsection{Validation and Released-Test Results}
\noindent\textbf{ALFWorld results.}
Under the same evolution and evaluation protocol, ALFWorld's three feedback runs generate 30 candidates, of which three establish new validation bests. Success-only produces the validation-selected round-9 skill and improves released-test performance by 2.2 points, whereas Normal is stronger on both robustness and transfer. Results appear in Tables~\ref{tab:detailed-scores}, \ref{tab:evolution-run-summary}, and~\ref{tab:task-scope-exact}.

\noindent\textbf{Identity-aware validation results.}
Gemini--OfficeQA Success-only evaluates the same skill eight times, with hard scores ranging from 71.43\% to 83.67\% (mean 76.28\%, standard deviation 3.92 points). Because artifact identity is unchanged, the branch remains at round 0. Gemini--DocVQA Success-only likewise retains the parent hash after its round-1 evaluation increase, and all three DeepSeek--OfficeQA views retain the parent. Their test and diagnostic entries characterize repeated execution of the parent artifact.

\begin{table}[H]
\centering
\footnotesize
\setlength{\tabcolsep}{2.3pt}
\renewcommand{\arraystretch}{1.10}
\begin{tabularx}{0.97\textwidth}{@{}L{0.15\textwidth} L{0.105\textwidth} C L{0.17\textwidth} C L{0.17\textwidth} C@{}}
\toprule
\textbf{Benchmark} & \textbf{Condition} & \textbf{Best $r$} & \multicolumn{2}{c}{\textbf{Validation}} & \multicolumn{2}{c}{\textbf{Released test}} \\
\cmidrule(lr){4-5}\cmidrule(lr){6-7}
& & & \textbf{Primary} & \textbf{Aux.} & \textbf{Primary} & \textbf{Aux.} \\
\midrule
\multirow{4}{*}{\textbf{SearchQA}}
 & Parent & 0 & 160/200 (80.00) & 88.00 & 1059/1400 (75.64) & 85.68 \\
 & Normal & 9 & 164/200 (82.00) & 88.78 & 1091/1400 (77.93) & 86.81 \\
 & Fail-only & 2 & 163/200 (81.50) & 88.52 & 1082/1400 (77.29) & 86.32 \\
 & Success-only & 4 & 163/200 (81.50) & 88.77 & 1085/1400 (77.50) & 86.84 \\
\midrule
\multirow{4}{*}{\textbf{OfficeQA}}
 & Parent & 0 & 38/49 (77.55) & 83.33 & 93/148 (62.84) & 67.72 \\
 & Normal & 9 & 43/49 (87.76) & 89.12 & 103/148 (69.59) & 70.43 \\
 & Fail-only & 8 & 42/49 (85.71) & 87.07 & 103/148 (69.59) & 70.24 \\
 & Success-only & 6 & 42/49 (85.71) & 86.39 & 99/148 (66.89) & 67.23 \\
\midrule
\multirow{4}{*}{\textbf{SpreadsheetBench}}
 & Parent & 0 & 16/39 (41.03) & 41.03 & 141/281 (50.18) & 50.18 \\
 & Normal & 3 & 32/39 (82.05) & 82.05 & 231/281 (82.21) & 82.21 \\
 & Fail-only & 6 & 32/39 (82.05) & 82.05 & 241/281 (85.77) & 85.77 \\
 & Success-only & 0 & 16/39 (41.03) & 41.03 & 134/281 (47.69) & 47.69 \\
\midrule
\multirow{4}{*}{\textbf{ALFWorld}}
 & Parent & 0 & 116/140 (82.86) & -- & 117/134 (87.31) & -- \\
 & Normal & 8 & 120/140 (85.71) & -- & 115/134 (85.82) & -- \\
 & Fail-only & 0 & 116/140 (82.86) & -- & 119/134 (88.81) & -- \\
 & Success-only & 9 & 121/140 (86.43) & -- & 120/134 (89.55) & -- \\
\midrule
\multirow{4}{*}{\textbf{LiveMath}}
 & Parent & 0 & 18/35 (51.43) & -- & 52/106 (49.06) & -- \\
 & Normal & 0 & 18/35 (51.43) & -- & 43/106 (40.57) & -- \\
 & Fail-only & 3 & 20/35 (57.14) & -- & 45/106 (42.45) & -- \\
 & Success-only & 0 & 18/35 (51.43) & -- & 46/106 (43.40) & -- \\
\midrule
\multirow{4}{*}{\textbf{DocVQA}}
 & Parent & 0 & 51/53 (96.23) & 97.81 & 344/374 (91.98) & 96.53 \\
 & Normal & 0 & 51/53 (96.23) & 97.81 & 343/374 (91.71) & 96.72 \\
 & Fail-only & 0 & 51/53 (96.23) & 97.81 & 343/374 (91.71) & 96.66 \\
 & Success-only & 0 & 51/53 (96.23) & 97.81 & 341/374 (91.18) & 96.24 \\
\bottomrule
\end{tabularx}
\normalsize
\caption{GPT-5.5 validation and released-test metrics for the parent and three feedback views across five primary benchmarks and ALFWorld.}
\label{tab:detailed-scores}
\vspace{0.35em}
\begin{minipage}{0.99\textwidth}
\footnotesize
\textit{Note.} Primary entries give successes over eligible $n$, followed by percentages in parentheses. Auxiliary entries are percentages under the metrics in Table~\ref{tab:benchmark-protocol}; -- denotes no distinct auxiliary metric. A run with best $r=0$ retains the parent skill, and its released-test entry is an independent execution of that artifact.
\end{minipage}
\end{table}

\begin{table}[H]
\centering
\footnotesize
\setlength{\tabcolsep}{2.2pt}
\renewcommand{\arraystretch}{1.00}
\begin{tabularx}{\textwidth}{@{}L{0.16\textwidth} L{0.14\textwidth} L{0.12\textwidth} C C C C C@{}}
\toprule
\textbf{Model} & \textbf{Benchmark} & \textbf{Condition} & \textbf{Best $r$} & \textbf{Val. primary} & \textbf{Val. aux.} & \textbf{Test primary} & \textbf{Test aux.} \\
\midrule
\multirow{20}{*}{Gemini 3.1 Pro}
& \multirow{4}{*}{SearchQA} & Parent & 0 & 79.00 & 85.97 & 76.93 & 86.07 \\
& & Normal & 1 & 82.00 & 88.42 & 76.57 & 85.71 \\
& & Fail-only & 6 & 81.50 & 88.80 & 77.57 & 86.66 \\
& & Success-only & 1 & 81.00 & 87.78 & 77.79 & 86.53 \\
\addlinespace[0.08em]
& \multirow{4}{*}{OfficeQA} & Parent & 0 & 77.55 & 79.25 & 65.54 & 66.55 \\
& & Normal$^{=P}$ & 0 & 77.55 & 79.25 & 65.54 & 66.10 \\
& & Fail-only$^{=P}$ & 0 & 77.55 & 79.25 & 64.86 & 65.70 \\
& & Success-only$^{=P}$ & 0 & 77.55 & 79.25 & 67.57 & 68.13 \\
\addlinespace[0.08em]
& \multirow{4}{*}{SpreadsheetBench} & Parent & 0 & 38.46 & -- & 41.99 & -- \\
& & Normal & 7 & 82.05 & -- & 79.72 & -- \\
& & Fail-only & 1 & 76.92 & -- & 81.49 & -- \\
& & Success-only & 4 & 48.72 & -- & 50.18 & -- \\
\addlinespace[0.08em]
& \multirow{4}{*}{LiveMath} & Parent & 0 & 42.86 & -- & 41.51 & -- \\
& & Normal & 9 & 74.29 & -- & 64.15 & -- \\
& & Fail-only & 10 & 74.29 & -- & 56.60 & -- \\
& & Success-only & 2 & 62.86 & -- & 61.32 & -- \\
\addlinespace[0.08em]
& \multirow{4}{*}{DocVQA} & Parent & 0 & 92.45 & 96.77 & 95.45 & 97.66 \\
& & Normal & 4 & 94.34 & 97.18 & 95.99 & 97.93 \\
& & Fail-only & 4 & 94.34 & 97.18 & 95.19 & 97.82 \\
& & Success-only$^{=P}$ & 0 & 92.45 & 96.77 & 95.72 & 98.14 \\
\midrule
\multirow{16}{*}{DeepSeek V4-Pro}
& \multirow{4}{*}{SearchQA} & Parent & 0 & 74.50 & 82.71 & 73.00 & 82.49 \\
& & Normal & 1 & 76.50 & 83.88 & 73.86 & 83.41 \\
& & Fail-only & 2 & 76.00 & 83.37 & 74.21 & 84.05 \\
& & Success-only$^{=P}$ & 0 & 74.50 & 82.71 & 72.00 & 82.14 \\
\addlinespace[0.08em]
& \multirow{4}{*}{OfficeQA} & Parent & 0 & 73.47 & 75.51 & 56.76 & 57.69 \\
& & Normal$^{=P}$ & 0 & 73.47 & 75.51 & 54.73 & 55.62 \\
& & Fail-only$^{=P}$ & 0 & 73.47 & 75.51 & 52.70 & 53.77 \\
& & Success-only$^{=P}$ & 0 & 73.47 & 75.51 & 56.76 & 57.76 \\
\addlinespace[0.08em]
& \multirow{4}{*}{SpreadsheetBench} & Parent & 0 & 38.46 & -- & 39.86 & -- \\
& & Normal & 3 & 76.92 & -- & 68.68 & -- \\
& & Fail-only & 1 & 69.23 & -- & 64.77 & -- \\
& & Success-only & 6 & 48.72 & -- & 39.15 & -- \\
\addlinespace[0.08em]
& \multirow{4}{*}{LiveMath} & Parent & 0 & 40.00 & -- & 10.38 & -- \\
& & Normal & 8 & 54.29 & -- & 20.75 & -- \\
& & Fail-only & 8 & 48.57 & -- & 28.30 & -- \\
& & Success-only$^{=P}$ & 0 & 40.00 & -- & 23.58 & -- \\
\bottomrule
\end{tabularx}
\normalsize
\caption{Gemini 3.1 Pro and DeepSeek V4-Pro validation and released-test metrics\protect\\for the parent and three feedback views across nine supported model--benchmark settings.}
\label{tab:crossmodel-detailed-scores}
\vspace{0.35em}
\begin{minipage}{0.99\textwidth}
\footnotesize
\textit{Note.} Scores are percentages under the primary and auxiliary metrics in Table~\ref{tab:benchmark-protocol}; -- denotes no distinct auxiliary metric. The superscript $^{=P}$ marks a feedback branch whose validation-best artifact is byte-identical to the parent. Its validation entries therefore match the parent; released-test entries report independent executions of the same artifact.
\end{minipage}
\end{table}

\subsection{Evolution Dynamics}

\begin{table}[H]
\centering
\footnotesize
\setlength{\tabcolsep}{2.5pt}
\renewcommand{\arraystretch}{0.96}
\begin{tabularx}{0.99\textwidth}{@{}L{0.14\textwidth} L{0.16\textwidth} L{0.10\textwidth} C C C C C@{}}
\toprule
\textbf{Model} & \textbf{Benchmark} & \textbf{View} & \shortstack{\textbf{Executed}\\\textbf{rounds}} & \shortstack{\textbf{First new}\\\textbf{best}} & \shortstack{\textbf{Best}\\$\boldsymbol{r}$} & \shortstack{\textbf{\# new}\\\textbf{bests}} & \shortstack{\textbf{Cross-view}\\\textbf{choice}} \\
\midrule
\multirow{15}{*}{GPT-5.5}
& \multirow{3}{*}{SearchQA} & Normal & 10 & 1 & 9 & 2 & Yes \\
& & Fail-only & 10 & 2 & 2 & 1 & -- \\
& & Success-only & 10 & 4 & 4 & 1 & -- \\
\addlinespace[0.1em]
& \multirow{3}{*}{OfficeQA} & Normal & 10 & 1 & 9 & 3 & Yes \\
& & Fail-only & 10 & 3 & 8 & 3 & -- \\
& & Success-only & 10 & 2 & 6 & 3 & -- \\
\addlinespace[0.1em]
& \multirow{3}{*}{SpreadsheetBench} & Normal & 8 & 2 & 3 & 2 & -- \\
& & Fail-only & 10 & 2 & 6 & 4 & Yes \\
& & Success-only & 6 & -- & 0 & 0 & -- \\
\addlinespace[0.1em]
& \multirow{3}{*}{LiveMath} & Normal & 5 & -- & 0 & 0 & -- \\
& & Fail-only & 8 & 3 & 3 & 1 & Yes \\
& & Success-only & 5 & -- & 0 & 0 & -- \\
\addlinespace[0.1em]
& \multirow{3}{*}{DocVQA} & Normal & 9 & -- & 0 & 0 & Parent \\
& & Fail-only & 7 & -- & 0 & 0 & -- \\
& & Success-only & 7 & -- & 0 & 0 & -- \\
\midrule
\multirow{15}{*}{Gemini 3.1 Pro}
& \multirow{3}{*}{SearchQA} & Normal & 10 & 1 & 1 & 1 & Yes \\
& & Fail-only & 10 & 5 & 6 & 2 & -- \\
& & Success-only & 10 & 1 & 1 & 1 & -- \\
\addlinespace[0.1em]
& \multirow{3}{*}{OfficeQA} & Normal & 10 & -- & 0 & 0 & Parent \\
& & Fail-only & 10 & -- & 0 & 0 & -- \\
& & Success-only & 10 & -- & 0 & 0 & -- \\
\addlinespace[0.1em]
& \multirow{3}{*}{SpreadsheetBench} & Normal & 10 & 1 & 7 & 4 & Yes \\
& & Fail-only & 10 & 1 & 1 & 1 & -- \\
& & Success-only & 10 & 2 & 4 & 2 & -- \\
\addlinespace[0.1em]
& \multirow{3}{*}{LiveMath} & Normal & 10 & 1 & 9 & 5 & Yes \\
& & Fail-only & 10 & 1 & 10 & 4 & -- \\
& & Success-only & 10 & 2 & 2 & 1 & -- \\
\addlinespace[0.1em]
& \multirow{3}{*}{DocVQA} & Normal & 10 & 4 & 4 & 1 & Yes \\
& & Fail-only & 10 & 4 & 4 & 1 & -- \\
& & Success-only & 10 & -- & 0 & 0 & -- \\
\midrule
\multirow{12}{*}{DeepSeek V4-Pro}
& \multirow{3}{*}{SearchQA} & Normal & 10 & 1 & 1 & 1 & Yes \\
& & Fail-only & 10 & 2 & 2 & 1 & -- \\
& & Success-only & 10 & -- & 0 & 0 & -- \\
\addlinespace[0.1em]
& \multirow{3}{*}{OfficeQA} & Normal & 10 & -- & 0 & 0 & Parent \\
& & Fail-only & 10 & -- & 0 & 0 & -- \\
& & Success-only & 10 & -- & 0 & 0 & -- \\
\addlinespace[0.1em]
& \multirow{3}{*}{SpreadsheetBench} & Normal & 10 & 1 & 3 & 3 & Yes \\
& & Fail-only & 10 & 1 & 1 & 1 & -- \\
& & Success-only & 10 & 2 & 6 & 3 & -- \\
\addlinespace[0.1em]
& \multirow{3}{*}{LiveMath} & Normal & 10 & 8 & 8 & 1 & Yes \\
& & Fail-only & 10 & 6 & 8 & 2 & -- \\
& & Success-only & 3 & -- & 0 & 0 & -- \\
\midrule
\multirow{3}{*}{GPT-5.5}
& \multirow{3}{*}{ALFWorld} & Normal & 10 & 8 & 8 & 1 & -- \\
& & Fail-only & 10 & -- & 0 & 0 & -- \\
& & Success-only & 10 & 6 & 9 & 2 & Yes \\
\bottomrule
\end{tabularx}
\normalsize
\caption{Evolution summary for the 42 primary feedback runs and three additional GPT-5.5--ALFWorld runs. The primary runs contain 388 evaluated candidates and 55 byte-distinct validation new bests. A dash under First new best and $r=0$ under Best $r$ indicate that the parent remains best; fewer than ten executed rounds indicate early stopping.}
\label{tab:evolution-run-summary}
\end{table}

\clearpage
\begin{figure}[H]
\centering
\includegraphics[width=\textwidth]{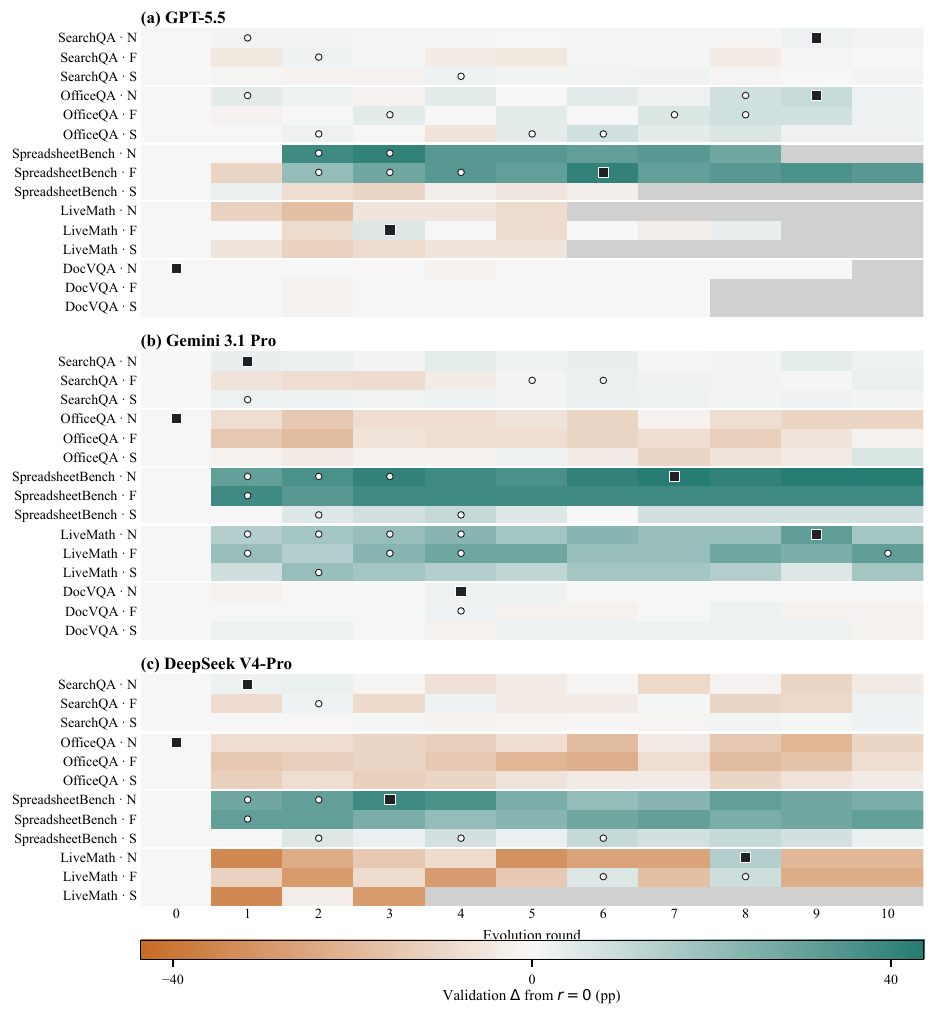}
\caption{Candidate validation differences from each setting's round-0 parent across the 42 primary feedback runs. Panels correspond to GPT-5.5, Gemini 3.1 Pro, and DeepSeek V4-Pro. White circles mark byte-distinct validation new bests, black squares mark setting-level selections, and gray cells mark rounds omitted after early stopping. Cells report raw validation differences; artifact improvements require a byte-distinct candidate. N, F, and S denote Normal, Fail-only, and Success-only.}
\label{fig:lifecycle-heatmap}
\end{figure}
\clearpage

\begin{table}[H]
\centering
\small
\setlength{\tabcolsep}{5.0pt}
\renewcommand{\arraystretch}{1.03}
\begin{tabularx}{0.84\textwidth}{@{}L{0.16\textwidth} *{10}{C}@{}}
\toprule
\textbf{Round} & 1 & 2 & 3 & 4 & 5 & 6 & 7 & 8 & 9 & 10 \\
\midrule
Candidates attempted & 42 & 42 & 42 & 41 & 41 & 39 & 38 & 36 & 34 & 33 \\
New-best events & 11 & 11 & 8 & 8 & 2 & 5 & 2 & 4 & 3 & 1 \\
Per-round yield (\%) & 26.2 & 26.2 & 19.0 & 19.5 & 4.9 & 12.8 & 5.3 & 11.1 & 8.8 & 3.0 \\
Cumulative events & 11 & 22 & 30 & 38 & 40 & 45 & 47 & 51 & 54 & 55 \\
\bottomrule
\end{tabularx}
\normalsize
\caption{Round-wise candidate yield across the 42 primary feedback runs. At each round, yield is the fraction of attempted candidates that establish a byte-distinct validation new best; the number of attempted candidates decreases as runs stop early.}
\label{tab:all-model-round-yield}
\end{table}

\begin{table}[H]
\centering
\small
\setlength{\tabcolsep}{5.0pt}
\renewcommand{\arraystretch}{1.05}
\begin{tabularx}{0.86\textwidth}{@{}L{0.20\textwidth} C C C C C C@{}}
\toprule
\textbf{Feedback evidence} & \textbf{Runs} & \textbf{Candidates} & \textbf{New bests} & \textbf{Yield} & \textbf{Runs improved} & \textbf{Selected} \\
\midrule
Normal & 14 & 132 & 23 & 17.4\% & 10/14 & 9 \\
Fail-only & 14 & 135 & 21 & 15.6\% & 11/14 & 2 \\
Failure-containing & 28 & 267 & 44 & 16.5\% & 21/28 & 11 \\
Success-only & 14 & 121 & 11 & 9.1\% & 6/14 & 0 \\
\bottomrule
\end{tabularx}
\normalsize
\caption{Search outcomes by feedback evidence across the 42 primary runs. The Failure-containing row aggregates Normal and Fail-only. Runs improved denotes at least one byte-distinct validation new best within a run; Selected counts setting-level evolved-skill selections.}
\label{tab:failure-bearing-search}
\end{table}

\noindent\textbf{Failure-containing feedback and late-round selection.}
Pooling Normal and Fail-only yields 44 byte-distinct validation new bests in 267 candidates (16.5\%), compared with 11/121 (9.1\%) for Success-only. At least one new best appears in 21/28 failure-containing runs (75.0\%) and 6/14 Success-only runs (42.9\%); all 11 primary evolved selections come from the former. Rounds 1--4 account for 38/55 new bests (69.1\%), while later rounds account for 17/55 (30.9\%). Yet six of the 11 selected evolved skills first appear in rounds 6--9, compared with five in rounds 1--4. Later discovery is less frequent but determines a majority of final evolved selections.

\subsection{Cross-Model SearchQA Results}

\begin{table}[H]
\centering
\small
\setlength{\tabcolsep}{3.8pt}
\renewcommand{\arraystretch}{1.08}
\begin{tabularx}{0.92\textwidth}{@{}L{0.23\textwidth} L{0.13\textwidth} C C C L{0.14\textwidth} C@{}}
\toprule
\textbf{Model} & \textbf{Selected view} & \textbf{Selected $r$} & \textbf{Val.\ $\Delta$} & \textbf{Test $\Delta$} & \textbf{Test-best view} & \textbf{Selection gap} \\
\midrule
Claude Opus 4.8 & Success-only & 7 & $+2.0$ & $+3.79$ & Success-only & $0.00$ \\
Kimi K2.5 & Fail-only & 2 & $+7.0$ & $+5.14$ & Success-only & $0.71$ \\
Gemini 3.1 Pro & Normal & 1 & $+3.0$ & $-0.36$ & Success-only & $1.21$ \\
DeepSeek V4-Pro & Normal & 1 & $+2.0$ & $+0.86$ & Fail-only & $0.36$ \\
GLM-5.1 & Fail-only & 5 & $+7.0$ & $+3.43$ & Normal & $0.07$ \\
Grok 4.5 & Normal & 3 & $+25.0$ & $+15.07$ & Fail-only & $0.36$ \\
Qwen3.5-Plus & Success-only & 3 & $+3.5$ & $+2.93$ & Fail-only & $0.57$ \\
\bottomrule
\end{tabularx}
\normalsize
\caption{SearchQA results for seven additional models. All three views execute ten rounds per model; among 210 candidates, 191 change the incoming skill and 29 establish a byte-distinct validation new best. Selection gap is the released-test score of the test-best feedback view minus that of the validation-selected view. GLM is evaluated on a common 1,398-item test subset, and Qwen3.5-Plus on 1,397 items. Differences are percentage points.}
\label{tab:cross-model-results}
\end{table}

\noindent\textbf{Eight-model SearchQA consistency.}
Including GPT-5.5, validation-selected skills improve released-test performance in seven of eight SearchQA models. Normal or Fail-only is selected in six models, while Success-only is selected for Claude Opus and Qwen3.5-Plus. Selected rounds span 1--9, extending the primary study's sparse, feedback-dependent dynamics across model families \cite{xie2026gala,xie2026agentsurvey,zhang2026setcon,zhang2025sec}.

\section{Post-Selection Generalization}
\label{app:post-selection-generalization}
\subsection{Deployment Variability}
Repeated evaluations cover the five GPT-5.5 benchmark settings, while robustness and transfer probes additionally include ALFWorld. Probe panels are fixed before evaluation and report \textit{N/A} when a tier is absent from the benchmark pool. The following tables give the probe-level results; Table~\ref{tab:selected-scope-crossmodel} gives aggregate robustness and transfer results for all selected primary skills.

\begin{table}[H]
\centering
\footnotesize
\renewcommand{\arraystretch}{0.97}
\setlength{\tabcolsep}{3pt}
\begin{tabularx}{0.99\textwidth}{L{0.15\textwidth} L{0.12\textwidth} C C C C C C C C}
\toprule
\textbf{Benchmark} & \textbf{Condition} & $\boldsymbol{n}$ & \textbf{Test $\Delta$} & \textbf{Repeat $\Delta_1$} & \textbf{Repeat $\Delta_2$} & \textbf{Repeat $\Delta_3$} & \textbf{Mean $\Delta$} & \textbf{Parent} & \textbf{Evolved} \\
\midrule
SearchQA & Normal & 100 & $+2.3$ & $-2.0$ & $-3.0$ & $0.0$ & $-1.7$ & 78.0 & 76.3 \\
OfficeQA & Normal & 57 & $+6.8$ & $+17.5$ & $+19.3$ & $+15.8$ & $+17.5$ & 58.5 & 76.0 \\
DocVQA & Parent skill & 100 & $-0.3$ & $-2.0$ & $-2.0$ & $+2.0$ & $-0.7$ & 90.7 & 90.0 \\
LiveMath & Fail-only & 100 & $-6.6$ & $+6.0$ & $+7.0$ & $+5.0$ & $+6.0$ & 52.3 & 58.3 \\
SpreadsheetBench & Normal & 100 & $+32.0$ & $+28.0$ & $+32.0$ & $+30.0$ & $+30.0$ & 54.0 & 84.0 \\
\bottomrule
\end{tabularx}
\normalsize
\caption{Three-repeat evaluation of GPT-5.5 artifacts on fixed benchmark-specific panels. Test $\Delta$ is the released-test difference; Repeat $\Delta_1$--$\Delta_3$ are paired-panel differences, and Parent and Evolved are their mean scores. The SpreadsheetBench row is the paired Normal diagnostic; Tables~\ref{tab:main-results} and~\ref{tab:task-scope-exact} report the selected Fail-only skill. DocVQA compares independent parent executions. Values are percentage points.}
\label{tab:deployment-variability}
\end{table}

\subsection{Robustness and Transfer}

\begin{table}[H]
\centering
\footnotesize
\renewcommand{\arraystretch}{0.97}
\setlength{\tabcolsep}{2.6pt}
\begin{tabularx}{0.99\textwidth}{L{0.15\textwidth} L{0.10\textwidth} >{\centering\arraybackslash}p{0.13\textwidth} C C C C C}
\toprule
& & \multicolumn{3}{c}{\textbf{Same-task robustness}} & \multicolumn{3}{c}{\textbf{Official-task transfer}} \\
\cmidrule(lr){3-5}\cmidrule(lr){6-8}
\textbf{Benchmark} & \textbf{Condition} & \textbf{R1 $\Delta$ ($n$)} & \textbf{R2 $\Delta$ ($n$)} & \textbf{R3 $\Delta$ ($n$)} & \textbf{T1 $\Delta$ ($n$)} & \textbf{T2 $\Delta$ ($n$)} & \textbf{T3 $\Delta$ ($n$)} \\
\midrule
\multirow{3}{*}{SearchQA} & Normal & $-0.7$ (97) & $+2.3$ (100) & $-1.0$ (100) & \textit{N/A} & $+7.3$ (100) & $+3.7$ (9) \\
 & Fail-only & $-2.1$ (97) & $+0.3$ (100) & $-1.3$ (100) & \textit{N/A} & $+8.3$ (100) & $0.0$ (9) \\
 & Success-only & $+1.0$ (97) & $+3.3$ (100) & $+1.3$ (100) & \textit{N/A} & $+6.3$ (100) & $0.0$ (9) \\
\addlinespace[0.2em]
\multirow{3}{*}{OfficeQA} & Normal & $+15.8$ (57) & $+10.0$ (50) & $+8.2$ (57) & $+20.0$ (5) & $+13.1$ (33) & $+2.1$ (47) \\
 & Fail-only & $+10.5$ (57) & $+5.3$ (50) & $+8.2$ (57) & $+20.0$ (5) & $+5.1$ (33) & $0.0$ (47) \\
 & Success-only & $+11.7$ (57) & $+1.3$ (50) & $+1.8$ (57) & $0.0$ (5) & $+10.1$ (33) & $+3.5$ (47) \\
\addlinespace[0.2em]
\multirow{3}{*}{ALFWorld} & Normal & $+3.0$ (44) & $+2.3$ (44) & $+1.5$ (44) & \textit{N/A} & $+2.3$ (44) & $0.0$ (11) \\
 & Fail-only$^\dagger$ & $+0.8$ (44) & $+2.3$ (44) & $-1.5$ (44) & \textit{N/A} & $-0.8$ (44) & $-3.0$ (11) \\
 & Success-only & $+3.0$ (44) & $+2.3$ (44) & $-1.5$ (44) & \textit{N/A} & $+3.0$ (44) & $-3.0$ (11) \\
\addlinespace[0.2em]
\multirow{3}{*}{SpreadsheetBench} & Normal & $+31.3$ (99) & $+36.3$ (100) & $+38.3$ (100) & \textit{N/A} & $+52.0$ (100) & $+3.5$ (57) \\
 & Fail-only & $+26.3$ (99) & $+39.0$ (100) & $+41.3$ (100) & \textit{N/A} & $+53.3$ (100) & $+2.9$ (57) \\
 & Success-only & $-10.1$ (99) & $+0.7$ (100) & $+0.7$ (100) & \textit{N/A} & $-1.0$ (100) & $+1.2$ (57) \\
\addlinespace[0.2em]
LiveMath & Fail-only & \shortstack{R1a: $+3.8$ (97)\\R1b: $+6.3$ (100)} & $+12.0$ (100) & $+11.0$ (100) & \textit{N/A} & $0.0$ (4) & $-33.3$ (2) \\
DocVQA & Parent skill & $-0.3$ (100) & $-2.0$ (100) & $-1.0$ (100) & $+1.9$ (18) & $-1.0$ (100) & $0.0$ (100) \\
\bottomrule
\end{tabularx}
\normalsize
\caption{GPT-5.5 probe-level robustness and transfer differences across five primary benchmarks and ALFWorld. Entries are three-deployment mean differences from parent; parentheses give eligible $n$. The table includes every validation-improving run and three parent-identical comparison conditions. Macro-averages weight available probes equally and omit \textit{N/A} tiers; LiveMath R1a and R1b count separately. Values are percentage points.}
\label{tab:task-scope-exact}
\end{table}

\begin{table}[H]
\centering
\footnotesize
\setlength{\tabcolsep}{3.0pt}
\renewcommand{\arraystretch}{1.06}
\begin{tabularx}{\textwidth}{@{}L{0.13\textwidth} L{0.14\textwidth} L{0.12\textwidth} >{\centering\arraybackslash}p{0.11\textwidth} *{7}{C}@{}}
\toprule
\textbf{Model} & \textbf{Benchmark} & \textbf{Selected skill} &
\textbf{R1 $\Delta$} & \textbf{R2 $\Delta$} & \textbf{R3 $\Delta$} &
\textbf{T1 $\Delta$} & \textbf{T2 $\Delta$} & \textbf{T3 $\Delta$} &
\textbf{R} & \textbf{T} \\
\midrule
GPT-5.5 & SearchQA & Normal, $r=9$ & $-0.7$ & $+2.3$ & $-1.0$ & \textit{N/A} & $+7.3$ & $+3.7$ & $+0.2$ & $+5.5$ \\
GPT-5.5 & OfficeQA & Normal, $r=9$ & $+15.8$ & $+10.0$ & $+8.2$ & $+20.0$ & $+13.1$ & $+2.1$ & $+11.3$ & $+11.7$ \\
GPT-5.5 & SpreadsheetBench & Fail-only, $r=6$ & $+26.3$ & $+39.0$ & $+41.3$ & \textit{N/A} & $+53.3$ & $+2.9$ & $+35.5$ & $+28.1$ \\
GPT-5.5 & LiveMath & Fail-only, $r=3$ & $+3.8/+6.3$ & $+12.0$ & $+11.0$ & \textit{N/A} & $0.0$ & $-33.3$ & $+8.3$ & $-16.7$ \\
\midrule
Gemini 3.1 Pro & SearchQA & Normal, $r=1$ & $-0.7$ & $+1.0$ & $-1.3$ & \textit{N/A} & $+3.0$ & $0.0$ & $-0.3$ & $+1.5$ \\
Gemini 3.1 Pro & SpreadsheetBench & Normal, $r=7$ & $+44.8$ & $+44.3$ & $+42.3$ & \textit{N/A} & $+50.3$ & $+0.6$ & $+43.8$ & $+25.5$ \\
Gemini 3.1 Pro & LiveMath & Normal, $r=9$ & $+23.7/+16.3$ & $+17.0$ & $+17.0$ & \textit{N/A} & $+16.7$ & $+16.7$ & $+18.5$ & $+16.7$ \\
Gemini 3.1 Pro & DocVQA & Normal, $r=4$ & $+1.0$ & $-0.7$ & $+0.7$ & $0.0$ & $-1.3$ & $+0.3$ & $+0.3$ & $-0.3$ \\
\midrule
DeepSeek V4-Pro & SearchQA & Normal, $r=1$ & $+1.4$ & $-0.7$ & $-1.3$ & \textit{N/A} & $+3.7$ & $0.0$ & $-0.2$ & $+1.8$ \\
DeepSeek V4-Pro & SpreadsheetBench & Normal, $r=3$ & $+33.7$ & $+31.0$ & $+32.3$ & \textit{N/A} & $+31.0$ & $+11.1$ & $+32.3$ & $+21.1$ \\
DeepSeek V4-Pro & LiveMath & Normal, $r=8$ & $+13.4/+7.7$ & $+11.0$ & $+0.7$ & \textit{N/A} & $+16.7$ & $+50.0$ & $+8.2$ & $+33.3$ \\
\bottomrule
\end{tabularx}
\normalsize
\caption{Probe-level robustness and transfer differences for all 11 byte-distinct skills selected in the primary study. R1 contains two panels for LiveMath and one for the other benchmarks. R and T are equal-weight macro-averages over available panels and match Table~\ref{tab:main-results}; panel definitions and sample sizes are in Tables~\ref{tab:probe-definitions} and~\ref{tab:probe-instantiations}. Values are percentage points relative to the corresponding parent.}
\label{tab:selected-scope-crossmodel}
\end{table}

\noindent\textbf{Breadth of downstream improvement.}
Among the 11 selected evolved skills, nine improve released-test performance, nine improve robustness, and nine improve transfer. Seven improve all three measures, and every selected skill improves at least one of robustness or transfer. SpreadsheetBench is the strongest cross-model case: all three models select an evolved skill, released-test gains range from 28.8 to 37.7 points, and robustness and transfer are positive for every model.
\par\smallskip
\noindent\textbf{Evaluation unit.}
Each model--benchmark--view cell retains a complete multi-round evolution trajectory. Fixed-panel scores and three post-selection robustness and transfer deployments characterize downstream behavior; cross-setting summaries aggregate the matched trajectories.

\subsection{Validation-to-Test Selection Gaps}

\begin{table}[H]
\centering
\small
\setlength{\tabcolsep}{3.8pt}
\renewcommand{\arraystretch}{1.03}
\begin{tabularx}{0.99\textwidth}{@{}L{0.17\textwidth} L{0.16\textwidth} L{0.15\textwidth} C L{0.15\textwidth} C >{\centering\arraybackslash}p{0.11\textwidth}@{}}
\toprule
\textbf{Model} & \textbf{Benchmark} & \textbf{Val.-selected} & \textbf{Test $\Delta$} & \textbf{Test-best} & \textbf{Test $\Delta$} & \textbf{Selection gap} \\
\midrule
GPT-5.5 & SearchQA & Normal & $+2.3$ & Normal & $+2.3$ & $0.0$ \\
GPT-5.5 & OfficeQA & Normal & $+6.8$ & Normal / Fail-only & $+6.8$ & $0.0$ \\
GPT-5.5 & SpreadsheetBench & Fail-only & $+35.6$ & Fail-only & $+35.6$ & $0.0$ \\
GPT-5.5 & LiveMath & Fail-only & $-6.6$ & Parent & $0.0$ & $6.6$ \\
GPT-5.5 & DocVQA & Parent & $0.0$ & Parent & $0.0$ & $0.0$ \\
\midrule
Gemini 3.1 Pro & SearchQA & Normal & $-0.4$ & Success-only & $+0.9$ & $1.3$ \\
Gemini 3.1 Pro & OfficeQA & Parent & $0.0$ & Parent & $0.0$ & $0.0$ \\
Gemini 3.1 Pro & SpreadsheetBench & Normal & $+37.7$ & Fail-only & $+39.5$ & $1.8$ \\
Gemini 3.1 Pro & LiveMath & Normal & $+22.6$ & Normal & $+22.6$ & $0.0$ \\
Gemini 3.1 Pro & DocVQA & Normal & $+0.5$ & Normal & $+0.5$ & $0.0$ \\
\midrule
DeepSeek V4-Pro & SearchQA & Normal & $+0.9$ & Fail-only & $+1.2$ & $0.3$ \\
DeepSeek V4-Pro & OfficeQA & Parent & $0.0$ & Parent & $0.0$ & $0.0$ \\
DeepSeek V4-Pro & SpreadsheetBench & Normal & $+28.8$ & Normal & $+28.8$ & $0.0$ \\
DeepSeek V4-Pro & LiveMath & Normal & $+10.4$ & Fail-only & $+17.9$ & $7.5$ \\
\bottomrule
\end{tabularx}
\normalsize
\caption{Released-test selection gap under validation-based cross-view selection in the 14 primary settings. The gap is the best released-test score among the parent and byte-distinct skills returned by the three feedback views minus that of the validation-selected skill. Parent-identical views share the parent entry. Values are percentage points.}
\label{tab:selection-gap}
\end{table}

\section{Artifact-Level Evidence}
\label{app:artifact-evidence}
\subsection{Selected Skills and Provenance}

\begin{table}[H]
\centering
\footnotesize
\setlength{\tabcolsep}{2.4pt}
\renewcommand{\arraystretch}{1.05}
\begin{tabularx}{0.99\textwidth}{@{}L{0.23\textwidth} L{0.13\textwidth} X L{0.17\textwidth}@{}}
\toprule
\textbf{Model--benchmark} & \textbf{Selected skill} & \textbf{Retained operational guidance} & \textbf{Observed scope} \\
\midrule
GPT-5.5--SearchQA & Normal, $r=9$ & Resolve the requested answer type; prefer canonical naming and concise answer-only output. & $+2.0/+2.3/+0.2/+5.5$ \\
GPT-5.5--OfficeQA & Normal, $r=9$ & Retrieve and read before computing; track operands and units; match the requested answer surface. & $+10.2/+6.8/+11.3/+11.7$ \\
GPT-5.5--SpreadsheetBench & Fail-only, $r=6$ & Inspect the workbook; execute a self-contained script; materialize values; save, reopen, and verify target cells. & $+41.0/+35.6/+35.5/+28.1$ \\
GPT-5.5--LiveMath & Fail-only, $r=3$ & Audit all options, quantifiers, hypotheses, equality cases, implication direction, and theorem strength; return only the label. & $+5.7/-6.6/+8.3/-16.7$ \\
\midrule
Gemini 3.1 Pro--SearchQA & Normal, $r=1$ & Remove unnecessary corporate suffixes; use common canonical names or surnames when appropriate; answer only. & $+3.0/-0.4/-0.3/+1.5$ \\
Gemini 3.1 Pro--SpreadsheetBench & Normal, $r=7$ & Use valid \texttt{openpyxl} syntax; handle duplicate headers and nulls; preserve blocks; compute values and verify with dual loading. & $+43.6/+37.7/+43.8/+25.5$ \\
Gemini 3.1 Pro--LiveMath & Normal, $r=9$ & Reject partial or overstrong options; inspect quantifiers and formulas; return the exact complete option text. & $+31.4/+22.6/+18.5/+16.7$ \\
Gemini 3.1 Pro--DocVQA & Normal, $r=4$ & Discriminate neighboring labels and preserve exact punctuation and symbols in the answer. & $+1.9/+0.5/+0.3/-0.3$ \\
\midrule
DeepSeek V4-Pro--SearchQA & Normal, $r=1$ & Identify the target entity type and return a concise canonical answer without elaboration. & $+2.0/+0.9/-0.2/+1.8$ \\
DeepSeek V4-Pro--SpreadsheetBench & Normal, $r=3$ & Separate sheet names; type-coerce comparisons; delete rows with marker-aware indexing; compute literal values and verify the saved workbook. & $+38.5/+28.8/+32.3/+21.1$ \\
DeepSeek V4-Pro--LiveMath & Normal, $r=8$ & Enforce the single-label contract and check meta-options and stronger-result traps before answering. & $+14.3/+10.4/+8.2/+33.3$ \\
\bottomrule
\end{tabularx}
\normalsize
\caption{Selected-skill change cards for the 11 byte-distinct evolved skills selected in the primary study.}
\label{tab:selected-skill-cards}
\vspace{0.35em}
\begin{minipage}{0.99\textwidth}
\footnotesize
\textit{Note.} The observed-scope entries report percentage-point differences in validation/released test/robustness/transfer order. Each row pairs a byte-distinct validation-selected artifact with its retained operational guidance and downstream scope.
\end{minipage}
\end{table}

\begin{table}[H]
\centering
\small
\setlength{\tabcolsep}{4.0pt}
\renewcommand{\arraystretch}{1.04}
\begin{tabularx}{0.99\textwidth}{@{}L{0.17\textwidth} L{0.17\textwidth} L{0.15\textwidth} L{0.16\textwidth} L{0.16\textwidth}@{}}
\toprule
\textbf{Model} & \textbf{Benchmark} & \textbf{Selected skill} & \textbf{Parent SHA-256} & \textbf{Selected SHA-256} \\
\midrule
GPT-5.5 & SearchQA & Normal, $r=9$ & \texttt{d3ed21de4a52} & \texttt{5f9c3ce70aef} \\
GPT-5.5 & OfficeQA & Normal, $r=9$ & \texttt{9d377d1c2906} & \texttt{ca689fecd40d} \\
GPT-5.5 & SpreadsheetBench & Fail-only, $r=6$ & \texttt{5c49d03008ae} & \texttt{859ab2927df6} \\
GPT-5.5 & LiveMath & Fail-only, $r=3$ & \texttt{84d545e168a0} & \texttt{e9bc725bfd2d} \\
\midrule
Gemini 3.1 Pro & SearchQA & Normal, $r=1$ & \texttt{d3ed21de4a52} & \texttt{5676bd5ff0ae} \\
Gemini 3.1 Pro & SpreadsheetBench & Normal, $r=7$ & \texttt{5c49d03008ae} & \texttt{26e7ec56b1fd} \\
Gemini 3.1 Pro & LiveMath & Normal, $r=9$ & \texttt{84d545e168a0} & \texttt{5ee24d6ca337} \\
Gemini 3.1 Pro & DocVQA & Normal, $r=4$ & \texttt{60fa9bae9059} & \texttt{5ce713b3c15f} \\
\midrule
DeepSeek V4-Pro & SearchQA & Normal, $r=1$ & \texttt{d3ed21de4a52} & \texttt{2b9b8273632c} \\
DeepSeek V4-Pro & SpreadsheetBench & Normal, $r=3$ & \texttt{5c49d03008ae} & \texttt{60478e37ca7e} \\
DeepSeek V4-Pro & LiveMath & Normal, $r=8$ & \texttt{84d545e168a0} & \texttt{be370bcfc14f} \\
\bottomrule
\end{tabularx}
\normalsize
\caption{Artifact identity for the 11 selected evolved skills. Twelve-character SHA-256 prefixes are shown for readability; every selected artifact differs from its parent.}
\label{tab:selected-skill-provenance}
\end{table}

\noindent\textbf{Validation selectivity and artifact identity.}
Validation filters 388 evaluated candidates to 55 byte-distinct new bests (14.2\%), and cross-view selection retains 11 evolved skills while keeping the parent in three settings. Each selected evolved skill has a hash distinct from its parent, linking the retained validation improvement to a persistent artifact update.

\subsection{Cross-Model Convergence in Selected Skills}
\label{app:cross-model-skill-convergence}

The selected artifacts reveal benchmark-level convergence across independently evolved skills \cite{su2026agentvista,zhang2025spftsql,zhang2026afttab,liu2025conceptnetargument}. The comparison below focuses on SearchQA, SpreadsheetBench, and LiveMath, for which all three primary models select byte-distinct evolved skills; Table~\ref{tab:selected-skill-cards} gives the corresponding setting-level cards.

\paragraph{SearchQA: answer-surface convergence.}
All three selected artifacts require concise canonical entity answers, explicit identification of the requested entity type, and answer-only output. GPT-5.5 Normal at round 9 further distinguishes property values from category nouns and creators from work titles. Gemini Normal at round 1 removes corporate suffixes and unnecessary name components, while DeepSeek Normal at round 1 makes entity-type and creator-versus-title resolution explicit.

\paragraph{SpreadsheetBench: executable postconditions.}
All three selected artifacts inspect the workbook, compute literal values in Python, preserve its structure, save the result, and verify the output. GPT-5.5 Fail-only at round 6 reopens the saved workbook with \texttt{data\_only=True} and verifies populated target cells. Gemini Normal at round 7 adds dual loading, duplicate-header handling, numeric-null rules, and sortable-block headers. DeepSeek Normal at round 3 adds sheet-name separation, type-coerced matching, and marker-aware row deletion.

\paragraph{LiveMath: theorem-level option comparison.}
All three selected artifacts compare options at theorem level, audit hypotheses and quantifiers, handle the recurring meta-option, and enforce the benchmark-specific answer contract. GPT-5.5 Fail-only at round 3 adds equality-case, implication-direction, and strongest-result checks. Gemini Normal at round 9 distinguishes partial from overstrong statements and returns the exact option text, while DeepSeek Normal at round 8 combines guarded meta-option selection with label-only output.

Across the nine artifacts, the shared retained mechanisms cluster by benchmark---answer normalization for SearchQA, executable workbook postconditions for SpreadsheetBench, and theorem-level option auditing for LiveMath---while model-specific specialization appears in the operational clauses.

\subsection{Comparative Case Studies of Feedback Views}
\label{app:feedback-case-studies}

We compare optimizer reports and \texttt{skill.md} differences from paired Normal and Fail-only runs. Each case connects the evidence visible to each view with retained rule changes and validation outcomes.

\paragraph{GPT-5.5--OfficeQA: evidence retrieval and answer format.}
The two views repeatedly encounter unsupported first-turn answers on numerical and table questions. In the Normal run, round 9 inspected 15 failed summaries and final trace turns, together with three passed traces for contrast. UID0101 answered an arc-elasticity question without retrieval and produced $-0.153$ rather than the expected $-1.162$; UID0010 answered a Treasury-value question from the prompt rather than the cited evidence; UID0086 and UID0160 added a percent sign when numeric-only answers were required. The accepted revision made targeted search and reading a first-action contract and specified the numeric percent surface. Fail-only round 8 saw the same unsupported-answer pattern without passed examples and retained a shorter hard-stop and search--read replacement rule. A subsequent Normal candidate at round 10 further expanded the checklist around the same failure pattern, but was rejected by validation.

\begin{table}[H]
\centering
\footnotesize
\setlength{\tabcolsep}{3.0pt}
\renewcommand{\arraystretch}{1.10}
\begin{tabularx}{\textwidth}{@{}L{0.09\textwidth} C L{0.22\textwidth} L{0.37\textwidth} C@{}}
\toprule
\textbf{View} & \textbf{Round} & \textbf{Evidence inspected} & \textbf{Retained or proposed \texttt{skill.md} change} & \textbf{Validation gate} \\
\midrule
Normal & 8 &
Read 16 failed samples and 3 passed contrasts; hard numerical failures frequently answered on turn 1 without evidence retrieval. &
Elevated retrieval to a pre-answer gate and added an ordered ledger for multi-stage transformations, units, and statistics. &
Accepted: 85.7 \\
Normal & 9 &
Read 15 failed summaries/final trace turns and 3 passed contrasts; unsupported first-turn answers remained dominant, with two numeric-percent surface failures. &
Required targeted search and reading before any answer, then specified when percent-change outputs omit \texttt{\%}. &
Accepted: 87.8 \\
Normal & 10 &
Read 8 of 17 failed samples and 2 passed contrasts; again found one-turn unsupported answers in FX, regression, and ``cannot be determined'' tasks. &
Proposed a literal stop condition and a detailed \texttt{grep $\rightarrow$ read $\rightarrow$ raw cells $\rightarrow$ compute} checklist, with expanded formatting rules. &
Rejected: 79.6 \\
Fail-only & 8 &
Read all 16 failed samples and no passed samples; every inspected failure answered on turn 1 without source retrieval. &
Added a hard stop against turn-1 answers and a concrete search--read--compute replacement pattern. &
Accepted: 85.7 \\
\bottomrule
\end{tabularx}
\normalsize
\caption{GPT-5.5--OfficeQA case study. Counts summarize the trajectories inspected by the optimizer; validation is the hard-score percentage on the fixed validation split.}
\label{tab:officeqa-feedback-case}
\end{table}

\paragraph{GPT-5.5--SpreadsheetBench: formula-like instructions and verifier-observable values.}
Both views encounter spreadsheet edits that write formula strings while the verifier reads \texttt{None} rather than the intended scalar values. Normal round 3 inspected 12 failed samples and two passed contrasts. It generalized the failure into broad formula-as-values, lookup, and full-grid population guidance. Fail-only round 6 inspected six failures without passed examples and focused on the verifier-visible condition: after saving, the output workbook must be reopened with \texttt{data\_only=True} and every required target cell must be populated. Both candidates reached 82.1 validation. The reported SpreadsheetBench result uses the round-6 Fail-only skill.

\begin{table}[H]
\centering
\footnotesize
\setlength{\tabcolsep}{3.2pt}
\renewcommand{\arraystretch}{1.10}
\begin{tabularx}{\textwidth}{@{}L{0.10\textwidth} C L{0.26\textwidth} L{0.36\textwidth} C@{}}
\toprule
\textbf{View} & \textbf{Round} & \textbf{Failure evidence} & \textbf{Accepted \texttt{skill.md} change} & \textbf{Validation gate} \\
\midrule
Normal & 3 &
Read 12 failed samples and 2 passed contrasts. Recurrent \texttt{cell\_level} tasks wrote formulas or incompletely filled copied ranges; the verifier read \texttt{None}. &
Added formula-as-values guidance for \texttt{INDEX/MATCH}, \texttt{SUMIFS}, \texttt{COUNTIFS}, and copied ranges, with a lookup-grid pattern that fills every target cell. &
Accepted: 82.1 \\
Fail-only & 6 &
Read all 6 failed samples. Formula-like tasks left target cells unpopulated or wrote formula strings that the verifier read as \texttt{None}. &
Made the verifier-visible requirement explicit: write scalar values, reopen the saved workbook with \texttt{data\_only=True}, and check that required target cells are not \texttt{None}. &
Accepted: 82.1 \\
\bottomrule
\end{tabularx}
\normalsize
\caption{GPT-5.5--SpreadsheetBench case study. The two views address the same verifier-observable failure but retain different levels of operational specificity.}
\label{tab:spreadsheet-feedback-case}
\end{table}

\paragraph{Success-only: stable specifications versus incidental patterns.}
The broader SearchQA runs provide two positive Success-only cases, while DeepSeek--LiveMath provides a contrasting rejected revision. These cases indicate that Success-only can help when positive trajectories repeatedly support a task-wide specification, whereas sparse positive evidence can support incidental rules. Table~\ref{tab:success-only-cases} summarizes the observed evidence and validation outcomes.

\begin{table}[H]
\centering
\footnotesize
\setlength{\tabcolsep}{3.2pt}
\renewcommand{\arraystretch}{1.08}
\begin{tabularx}{\textwidth}{@{}L{0.20\textwidth} C L{0.22\textwidth} L{0.34\textwidth} C C@{}}
\toprule
\textbf{Setting} & \textbf{Round} & \textbf{Observed positive evidence} & \textbf{Candidate rule} & \textbf{Val.\ gate} & \textbf{Test $\Delta$} \\
\midrule
DeepSeek--LiveMath & 3 &
Eight successful trajectories in the round sample. &
Added a ``strongest/equivalence'' heuristic and changed the required output from an option label to full option text. &
$40.0\rightarrow11.4$; rejected & -- \\
Claude Opus--SearchQA & 7 &
Successful traces repeatedly identify the clue referent and use a short canonical answer. &
Made referent identification and concise canonical answer form explicit. &
$77.0\rightarrow79.0$; selected & $+3.79$ \\
Qwen3.5-Plus--SearchQA & 3 &
Successful traces support a shared answer-form specification across questions. &
Retained the common short-answer specification without adding task-specific content. &
$73.0\rightarrow76.5$; selected & $+2.93$ \\
\bottomrule
\end{tabularx}
\normalsize
\caption{Contrasting Success-only cases. The LiveMath candidate is rejected and the branch retains its parent; the two validation-selected SearchQA skills encode shared answer-form specifications supported across successful traces.}
\label{tab:success-only-cases}
\end{table}

\subsection{End-to-End Evolution Traces}
\label{app:end-to-end-traces}

Tables~\ref{tab:trace-gpt-searchqa}--\ref{tab:trace-deepseek-livemath} connect score changes to candidate content across three complete branches: a late answer-form improvement, a cumulative procedural repair, and a mixed-versus-positive-only contrast. ``Feedback used'' reports the trajectories visible to each optimizer call; validation scores and gate actions follow the recorded evolution histories.

\paragraph{Late answer-form repair.}
GPT-5.5--SearchQA first improves in round 1, when a broad minimal-span rule raises validation from 80.0 to 81.0. Rounds 2--8 repeatedly revise how creator names and surnames should be expressed, but none surpasses that checkpoint. Round 9 then makes a narrower change---return the value of a requested property without its category noun---and establishes the selected 82.0 checkpoint. The late gain is a one-clause correction discovered after several plausible alternatives fail validation.

\paragraph{Procedural repair.}
The Gemini SpreadsheetBench run improves from 38.5 to 69.2 in round 1 by replacing unevaluated formulas with literal values. Rounds 2 and 3 add import-path and dual-load safeguards, each contributing another 5.1 points. Three subsequent proposals fail to improve the best. Round 7 then addresses four concrete execution defects---cell-call syntax, duplicate headers, numeric nulls, and headers inside sortable blocks---and raises validation to 82.1. Later revisions tie or regress and are rolled back.

\paragraph{Negative evidence and validation gating.}
DeepSeek-V4-Pro--LiveMath illustrates the role of rejected candidates in the evolution trajectory. Normal proposes seven unsuccessful revisions before the round-8 combination of label formatting and guarded meta-option selection improves validation from 40.0 to 54.3; the two later candidates fall to 20.0 and do not replace it. Success-only never exceeds the parent. Its round-3 revision infers a strength/equivalence rule from eight successes and changes the output contract to full option text, reducing validation to 11.4. The branch then has no eligible positive feedback and stops with the parent intact.

\begin{table}[H]
\centering
\footnotesize
\setlength{\tabcolsep}{3.0pt}
\renewcommand{\arraystretch}{1.05}
\begin{tabularx}{\textwidth}{@{}C L{0.12\textwidth} L{0.49\textwidth} C L{0.11\textwidth} C@{}}
\toprule
\textbf{r} & \textbf{Feedback used} & \textbf{Candidate revision} & \textbf{Val.} & \textbf{Gate} & \textbf{Best r} \\
\midrule
0 & -- & Round-0 parent; no learned rules. & 80.0 & Baseline & 0 \\
1 & 34 S + 6 F & Add minimal answer spans, surname and company shortening, and creator-relation checks. & 81.0 & New best & 1 \\
2 & 39 S + 1 F & Prefer the creator for bare descriptions of books, films, series, or trilogies. & 81.0 & Flat & 1 \\
3 & 39 S + 1 F & Narrow the creator cue to descriptions such as ``a trilogy/series set in \ldots''. & 79.0 & Reject & 1 \\
4 & 39 S + 1 F & Broaden the creator cue to bare ``set in'' and ``about'' descriptions. & 80.5 & Flat & 1 \\
5 & 38 S + 2 F & Balance surname-only answers with full creator names when the clue does not identify the person. & 80.0 & Flat & 1 \\
6 & 40 S + 0 F & No byte-distinct candidate (\texttt{VALID\_NULL}). & 79.0 & Reject & 1 \\
7 & 37 S + 3 F & Make surname-only the default for biographical clues unless the surname is ambiguous. & 79.0 & Reject & 1 \\
8 & 38 S + 2 F & Use full names for ambiguous office, title, role, or family-relation clues. & 79.5 & Flat & 1 \\
9 & 39 S + 1 F & For a property query, return only the value word rather than the category noun. & 82.0 & New best & 9 \\
10 & 40 S + 0 F & No byte-distinct candidate (\texttt{VALID\_NULL}). & 81.0 & Flat & 9 \\
\bottomrule
\end{tabularx}
\normalsize
\caption{Complete GPT-5.5--SearchQA Normal evolution. Two of ten proposals establish byte-distinct validation bests; the selected skill first appears in round 9.}
\label{tab:trace-gpt-searchqa}
\end{table}

\begin{table}[H]
\centering
\footnotesize
\setlength{\tabcolsep}{3.0pt}
\renewcommand{\arraystretch}{1.05}
\begin{tabularx}{\textwidth}{@{}C L{0.12\textwidth} L{0.49\textwidth} C L{0.11\textwidth} C@{}}
\toprule
\textbf{r} & \textbf{Feedback used} & \textbf{Candidate revision} & \textbf{Val.} & \textbf{Gate} & \textbf{Best r} \\
\midrule
0 & -- & Round-0 parent. & 38.5 & Baseline & 0 \\
1 & 16 S + 24 F & Compute literal values in Python; avoid unevaluated Excel formulas and unnecessary \texttt{pandas}. & 69.2 & New best & 1 \\
2 & 34 S + 6 F & Add the sandbox-safe import-path workaround before loading \texttt{openpyxl}. & 74.4 & New best & 2 \\
3 & 33 S + 7 F & Add dual workbook loading for formula values and reverse-order row deletion. & 79.5 & New best & 3 \\
4 & 33 S + 7 F & Emulate blank-as-zero formula behavior, delete rows one by one, and exclude headers from sorting. & 76.9 & Reject & 3 \\
5 & 33 S + 7 F & Add date/time conversion and an explore-first workbook inspection procedure. & 74.4 & Reject & 3 \\
6 & 33 S + 7 F & Ban \texttt{pandas} and use a two-pass row-deletion procedure. & 79.5 & Flat & 3 \\
7 & 33 S + 7 F & Fix cell-call syntax, duplicate-header mapping, numeric nulls, and headers inside contiguous blocks. & 82.1 & New best & 7 \\
8 & 35 S + 5 F & Add dynamic header discovery, date-type checks, and safeguards for relative deletions. & 79.5 & Reject & 7 \\
9 & 34 S + 6 F & Add partial header matching, occurrence-aligned duplicate mapping, and rounded time conversion. & 82.1 & Flat & 7 \\
10 & 34 S + 6 F & Use \texttt{zip} for duplicate columns and explicit zero defaults for numeric grids. & 82.1 & Flat & 7 \\
\bottomrule
\end{tabularx}
\normalsize
\caption{Complete Gemini-3.1-Pro--SpreadsheetBench Normal evolution. Validation retains four cumulative procedural revisions and rolls back the other six.}
\label{tab:trace-gemini-spreadsheet}
\end{table}

\begin{table}[H]
\centering
\footnotesize
\setlength{\tabcolsep}{2.8pt}
\renewcommand{\arraystretch}{1.04}
\begin{tabularx}{\textwidth}{@{}L{0.10\textwidth} C L{0.10\textwidth} L{0.43\textwidth} C L{0.10\textwidth} C@{}}
\toprule
\textbf{View} & \textbf{r} & \textbf{Feedback used} & \textbf{Candidate revision} & \textbf{Val.} & \textbf{Gate} & \textbf{Best r} \\
\midrule
Normal & 0 & -- & Round-0 parent. & 40.0 & Baseline & 0 \\
Normal & 1 & 3 S + 33 F & Add a best-effort, nonempty fallback. & 5.7 & Reject & 0 \\
Normal & 2 & 3 S + 33 F & Require a single-label guess when uncertain. & 17.1 & Reject & 0 \\
Normal & 3 & 5 S + 31 F & Add strength-aware comparison among options. & 25.7 & Reject & 0 \\
Normal & 4 & 3 S + 33 F & Require a valid A--E label and add a recurring meta-option rule. & 31.4 & Reject & 0 \\
Normal & 5 & 3 S + 33 F & Require a nonempty label for the strongest remaining option. & 8.6 & Reject & 0 \\
Normal & 6 & 3 S + 33 F & Expand the stronger-result and meta-option heuristic. & 14.3 & Reject & 0 \\
Normal & 7 & 2 S + 34 F & Recognize the recurring ``one remaining option'' meta-option. & 14.3 & Reject & 0 \\
Normal & 8 & 3 S + 33 F & Combine exact label formatting with a guarded meta-option strategy. & 54.3 & New best & 8 \\
Normal & 9 & 11 S + 25 F & Separate label-versus-statement output and add broader theorem checks. & 20.0 & Reject & 8 \\
Normal & 10 & 4 S + 32 F & Make the meta-option default more aggressive while retaining label-only output. & 20.0 & Reject & 8 \\
\midrule
Success-only & 0 & -- & Round-0 parent. & 40.0 & Baseline & 0 \\
Success-only & 1 & 3 S & Add no task rule; the generated diagnostic report changes the candidate bytes. & 5.7 & Reject & 0 \\
Success-only & 2 & 5 S & Add a ``strongest statement'' heuristic. & 37.1 & Reject & 0 \\
Success-only & 3 & 8 S & Add strength/equivalence rules and change the required output from a label to full option text. & 11.4 & Reject & 0 \\
Success-only & stop & 0 S & No eligible positive feedback remains after round 3; retain the parent without another optimizer call. & -- & Early stop & 0 \\
\bottomrule
\end{tabularx}
\normalsize
\caption{Complete DeepSeek-V4-Pro--LiveMath trajectory contrast. Normal selects its round-8 repair after seven candidate revisions; Success-only evaluates three revisions before the positive-feedback pool is exhausted.}
\label{tab:trace-deepseek-livemath}
\end{table}

\subsection{Representative Selected-Skill Excerpts}
\label{app:verbatim-skills}

The selected skills encode three distinct forms of retained revision: answer-surface normalization for SearchQA, theorem-level option comparison for LiveMath, and an executable workbook workflow for SpreadsheetBench.

\noindent\begin{minipage}{\textwidth}
\begin{lstlisting}[
  style=skillartifact,
  basicstyle=\ttfamily\footnotesize,
  caption={GPT-5.5--SearchQA, Normal, round 9.},
  label={lst:gpt-searchqa-skill},
  captionpos=b
]
# Question Answering Skill

Answer with the shortest conventional answer that satisfies the clue.

- Prefer the common quiz-bowl/search answer form over a fuller encyclopedia name: for people, use surname-only when that is the conventional unambiguous answer, especially for artists, authors, or military figures in biographical clues anchored by dates, deaths, places, or famous works; use the full conventional name when the clue identifies the person mainly by office, title, role, or family relation that could fit multiple people with the same surname. Omit corporate suffixes such as "Corporation" when the common company name is enough, and reduce descriptive phrases to the requested property when the clue asks for "this kind/type of" something; for a property value such as a charge, color, direction, or polarity, give only the value word, not the category noun.
- Read the clue's relation before choosing the span. If the clue gives a title, subtitle, setting, or work description, ask whether it wants the creator/author rather than another part of the title or the series name; for bare descriptions of a book, film, series, or trilogy, including clues phrased only as "a trilogy/book/film/series set in..." or "about...", prefer the creator/author unless the wording explicitly asks for the work's name, and give the creator's full conventional name rather than surname-only when the clue does not already name them.
- Do not pad the answer with explanatory context. Return only the minimal entity, person, place, property, or title needed for an exact match.
\end{lstlisting}

\begin{lstlisting}[
  style=skillartifact,
  basicstyle=\ttfamily\footnotesize,
  caption={DeepSeek-V4-Pro--LiveMath, Normal, round 8.},
  label={lst:deepseek-livemath-skill},
  captionpos=b
]
# Live Mathematical MCQ Heuristics

## Answer-Format Trap
- When the expected answer is a single option label (A, B, C, D, E), output **exactly** that character with no surrounding text, no explanation, no Markdown, and no LaTeX. Never emit an empty string, a partial LaTeX fragment, or a quoted value. If the answer is `E`, output `E` alone.

## Option Selection Strategy
- Most livemath questions have a hidden meta-option: one of the listed options is correct, but a stronger result can be proven. If a question asks for "the strongest statement that can be proved," check whether any option claims a maximal or exhaustive classification when only a partial result is available. The canonical correct choice in many prompts is `One of the remaining options is correct, but a stronger result can be proven`.

## Theorem-Level Precision
- Check whether an option weakens the conclusion by dropping a characterization, equality clause, or full equivalence.
- Check whether an option overstates the theorem by upgrading regularity, removing scale restrictions, or changing an existential statement into a universal one.

## Hypotheses
- Verify the hypotheses and domain carefully. Distractors often keep the theorem shape but alter the required assumptions.
- Pay close attention to equality cases, extremal conditions, and whether a result applies to the full family or only a restricted subfamily.

## Final Answer
- Output the final answer as the single option label only.
\end{lstlisting}
\end{minipage}

\clearpage
\refstepcounter{lstlisting}
\label{lst:gemini-spreadsheet-skill}
\noindent\begin{minipage}[t]{0.485\textwidth}
{\footnotesize\bfseries Part I: lines 30--86\par}
\begin{lstlisting}[
  style=skillartifact,
  firstline=30,
  lastline=86,
  firstnumber=30,
  numbers=left,
  numberstyle=\tiny,
  stepnumber=5,
  numbersep=5pt,
]
# Spreadsheet Manipulation Skill (xlsx)

## Overview
This skill guides agents in manipulating Excel (.xlsx) spreadsheets using Python.

**Primary libraries**: `openpyxl` (structure-preserving read/write). `pandas` can be used for data transformation, but is sometimes unavailable in the environment.
Never use any other third-party libraries.

---

## Environment Sandbox Workaround (CRITICAL)

**CRITICAL:** The evaluation environment has a strict security sandbox that blocks `os.listdir('/tmp')`. Because your script executes from `/tmp`, standard library imports triggered by `openpyxl` (such as `numpy`, `PIL`, or `datetime`) will frequently raise a `RuntimeError: SKILLOPT_GENERATED_CODE_FILE_SCOPE_BLOCK: os.listdir:/tmp`. 

To prevent this crash and save execution turns, you **must** filter `/tmp` out of `sys.path` at the very beginning of your script, *before* importing `openpyxl`.

---

## Python Syntax Pitfalls (CRITICAL)

Always use explicitly named keyword arguments for both row and column when accessing cells: `ws.cell(row=r, column=c)`. Using a positional argument after a keyword argument (e.g., `ws.cell(row=r, c)`) causes a fatal `SyntaxError` and immediately fails the task.

---

## Python Computation vs. Excel Formulas

**CRITICAL:** Even if the instruction asks to "create an Excel formula" (e.g., "create a formula to sum...", "write a VLOOKUP..."), you **MUST compute the actual values in Python** and write the literal results back to the cells. 

Do **not** write Excel formula strings (like `=VLOOKUP(...)`, `=SUM(...)`, `=IF(...)`) into the workbook. The evaluation environment evaluates cell values directly; if you write a formula string, `openpyxl` cannot evaluate it, and the tests will see `None` or the literal formula string, causing a failure.

**Bad (Will Fail):**
```python
# Do not do this - the formula will not be evaluated!
ws.cell(row=2, column=3).value = f"=A2+B2"
ws.cell(row=3, column=3).value = f"=VLOOKUP(D3, 'Data'!A:B, 2, FALSE)"
```

**Good (Do This):**
```python
# Compute the result in Python and write the literal value
val_a = ws_data.cell(row=2, column=1).value or 0
val_b = ws_data.cell(row=2, column=2).value or 0
ws.cell(row=2, column=3).value = val_a + val_b
```

---

## The Dual-Load Pattern (CRITICAL for Reading Formulas)

When reading from a workbook that contains formulas, `openpyxl` will return the literal formula string (e.g., `"=A1+B1"`) instead of the computed value. To correctly read computed values while preserving the workbook's formatting and formulas when saving, you must load the workbook **twice**:

1. Load with `data_only=True` to **read** computed values.
2. Load standard (without `data_only`) to **write** and **save**.

**CRITICAL:** Never call `.save()` on the `data_only=True` workbook, as it will silently destroy all formulas in the file. Always write your results to the standard workbook and save that one.

---
\end{lstlisting}
\end{minipage}\hfill
\begin{minipage}[t]{0.485\textwidth}
{\footnotesize\bfseries Part II: lines 87--140\par}
\begin{lstlisting}[
  style=skillartifact,
  firstline=87,
  lastline=140,
  firstnumber=87,
  numbers=left,
  numberstyle=\tiny,
  stepnumber=5,
  numbersep=5pt,
]

## Text Matching & Duplicate Headers

When searching for sheet names, headers, or specific text, **always use case-insensitive matching** and strip whitespace. Prompt instructions frequently differ in casing from the actual workbook data.

Additionally, spreadsheets often contain multiple columns with the exact same header. If you map headers using a simple dictionary (`header_map[name] = col`), you will overwrite previous columns and map data incorrectly. Use a list to store all column indices for a given header:

```python
headers = {}
for c in range(1, ws.max_column + 1):
    val = ws.cell(row=1, column=c).value
    if val:
        name = str(val).strip().lower()
        headers.setdefault(name, []).append(c)
```

---

## Nulls vs. Zeros

When a task involves calculations, aggregating data, or filling a grid, be careful with empty cells. If an expected result cell represents a count, sum, or financial zero, write `0` instead of leaving the cell empty (`None`). Leaving cells as `None` when a numerical `0` is expected will cause evaluation failures.

---

## Safely Deleting Rows

When deleting rows based on a condition, always iterate over the row indices in **reverse order (descending)**. Otherwise, deleting a row shifts the remaining rows up, changing their indices, which causes you to skip rows or delete the wrong ones.

```python
# 1. Find all rows to delete (using ws_data to evaluate conditions)
rows_to_delete = []
for r in range(1, ws_data.max_row + 1):
    if ws_data.cell(row=r, column=1).value == "DELETE":
        rows_to_delete.append(r)

# 2. Sort descending and delete one by one from the standard workbook
for r in sorted(rows_to_delete, reverse=True):
    ws.delete_rows(r)
```

---

## Contiguous Blocks and Headers

When a task asks you to process or sort contiguous blocks of data separated by blank rows, ensure you do not mistakenly include the header row of each block in your sort. Headers should remain at the top of their respective blocks.

---

## Common Workflow

1. **Explore** the input file: list sheets, inspect headers, check dimensions.
2. **Write `solution.py`** with `INPUT_PATH` and `OUTPUT_PATH` defined at the top, utilizing the Dual-Load pattern.
3. **Execute** `python solution.py` and verify the output file was created.
4. **Confirm** the target cells/range contain the expected literal values (not formulas).
\end{lstlisting}
\end{minipage}

{\small\centering Listing~\thelstlisting: Gemini-3.1-Pro--SpreadsheetBench, Normal, round 7 (operational excerpt).\par}

\clearpage
\section{Test-Time Scaling and Verifier Diagnostics}
\label{app:tts-diagnostics}
\subsection{Test-Time Scaling Calculation}
\label{app:tts-calculation}

\begin{table}[H]
\centering
\small
\setlength{\tabcolsep}{4pt}
\begin{tabularx}{0.78\textwidth}{L{0.25\textwidth} C C C C}
\toprule
\textbf{Benchmark} & \textbf{Evolved} & \textbf{Parallel} & \textbf{Sequential} & \textbf{Gap} \\
\midrule
SearchQA & $+2.29$ & $+1.86$ & $+0.14$ & $0.43$ \\
SpreadsheetBench & $+35.23$ & $+4.27$ & $-5.34$ & $30.96$ \\
\bottomrule
\end{tabularx}
\normalsize
\caption{Primary test-time-scaling differences at maximum budget. Evolved, Parallel, and Sequential report percentage-point changes from the parent; Gap is Evolved minus Parallel.}
\label{tab:tts-differences}
\end{table}

For item $i$, let $z_{ik}\in\{0,1\}$ be the frozen verifier outcome of attempt $k$, and let $D_k$ be the outcome-independent set of items assigned a $k$th attempt, with $D_{k+1}\subseteq D_k$ and $D_1$ equal to the evaluation panel. On common support $D_K$, Parallel Sampling uses oracle any-success, whereas Sequential Refinement uses only the last response:
\[
\begin{aligned}
S_{\mathrm{par}}(K)
  &= \frac{100}{|D_K|}\sum_{i\in D_K}\max_{1\leq k\leq K} z_{ik}, \\
S_{\mathrm{seq}}(K)
  &= \frac{100}{|D_K|}\sum_{i\in D_K} z_{iK}.
\end{aligned}
\]
For the item-specific allocation $K_i$ used by the full SearchQA panel, the corresponding dynamic scores are
\[
S_{\mathrm{par}}^{\mathrm{dyn}}
=\frac{100}{n}\sum_{i=1}^{n}\max_{1\leq k\leq K_i}z_{ik},\qquad
S_{\mathrm{seq}}^{\mathrm{dyn}}
=\frac{100}{n}\sum_{i=1}^{n}z_{iK_i}.
\]
Let $z_i^{\mathrm{parent}}$ and $z_i^{\mathrm{evo}}$ be the frozen verifier outcomes of one-call parent and evolved-skill execution. Their scores are
\[
S_{\mathrm{parent}}=\frac{100}{n}\sum_i z_i^{\mathrm{parent}},
\qquad
S_{\mathrm{evo}}=\frac{100}{n}\sum_i z_i^{\mathrm{evo}}.
\]
For $m\in\{\mathrm{par},\mathrm{seq},\mathrm{evo}\}$, we report changes from the one-call parent as $\Delta_m=S_m-S_{\mathrm{parent}}$. Because attempt 1 reuses the frozen parent output, the number of \emph{new} target calls for one control is
\[
C_{\mathrm{add}}=\sum_{k=2}^{K_{\max}}|D_k|,
\qquad
C_{\mathrm{all}}=|D_1|+C_{\mathrm{add}},
\]
where $C_{\mathrm{all}}$ counts all score-bearing attempts, including the reused baseline. For fixed-panel controls, $C_{\mathrm{add}}=n(K_{\max}-1)$. Parallel oracle any-success assumes perfect post-hoc selection and is therefore an upper bound, not a deployable pass@1 estimator.

\begin{figure}[H]
\centering
\includegraphics[width=0.94\textwidth]{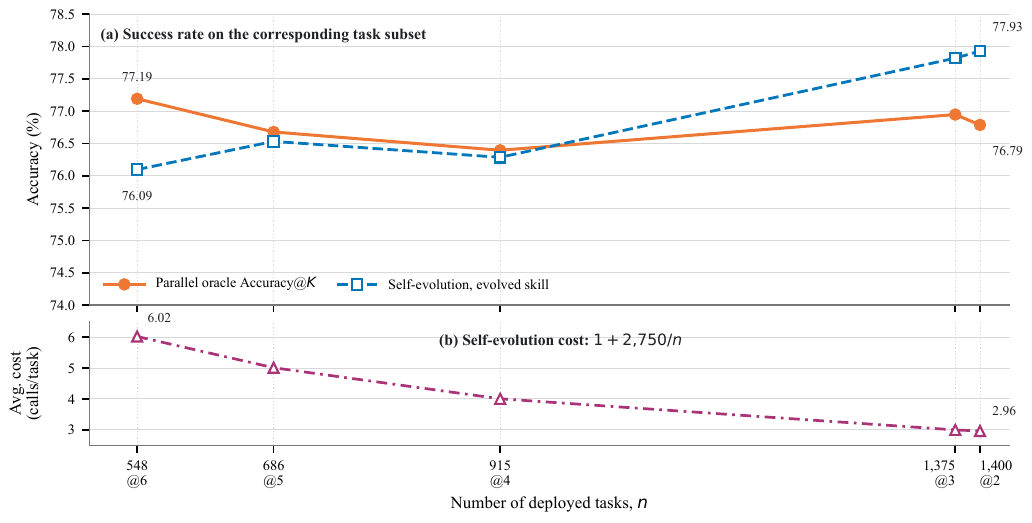}
\caption{GPT-5.5 SearchQA accuracy and amortized evolution cost across deployment scales. Each x tick pairs deployment size $n$ with Parallel budget $K$ on the corresponding nested subset. The upper panel compares oracle Accuracy@$K$ with one-call evolved-skill deployment; the lower panel reports $1+2{,}750/n$ calls per deployment for the selected evolution run. Parallel uses 2,740--2,750 additional calls at $n\in\{548,686,915,1375\}$ and 1,400 calls at the full-panel $K=2$ point.}
\label{fig:searchqa-tts-amortization}
\end{figure}

For Figure~\ref{fig:searchqa-tts-amortization}, $B_{\mathrm{evo}}=2{,}750$ counts target-model calls in the selected SearchQA Normal evolution run. Its amortized selected-run call metric over $n$ deployments is $C_{\mathrm{avg}}(n)=1+B_{\mathrm{evo}}/n$, where one is the evolved-skill deployment call. This metric covers the selected evolution run and subsequent deployments.

\begin{table}[H]
\centering
\footnotesize
\setlength{\tabcolsep}{4pt}
\begin{tabularx}{0.99\textwidth}{L{0.17\textwidth} C C L{0.29\textwidth} C C}
\toprule
\textbf{Benchmark} & $\boldsymbol{n}$ & $\boldsymbol{K_{\max}}$ & $\boldsymbol{(|D_2|,\ldots,|D_{K_{\max}}|)}$ & $\boldsymbol{C_{\mathrm{add}}}$ & $\boldsymbol{C_{\mathrm{all}}}$ \\
\midrule
SearchQA & 1,400 & 6 & $(1{,}400,1{,}375,915,686,548)$ & 4,924 & 6,324 \\
OfficeQA & 148 & 6 & $(148,148,148,148,148)$ & 740 & 888 \\
DocVQA & 374 & 4 & $(374,374,374)$ & 1,122 & 1,496 \\
LiveMath & 106 & 8 & $(106,106,106,106,106,106,106)$ & 742 & 848 \\
SpreadsheetBench & 281 & 4 & $(281,281,281)$ & 843 & 1,124 \\
\bottomrule
\end{tabularx}
\normalsize
\caption{GPT-5.5 test-time-scaling allocations and call counts. SearchQA uses a nested, outcome-independent allocation: 25 items receive $K_i=2$, 460 receive 3, 229 receive 4, 138 receive 5, and 548 receive 6 attempts. The other benchmarks evaluate every item through $K_{\max}$, with identical allocations for Parallel and Sequential.}
\label{tab:tts-definition-budget}
\end{table}

\begin{figure}[H]
\centering
\includegraphics[width=0.94\textwidth]{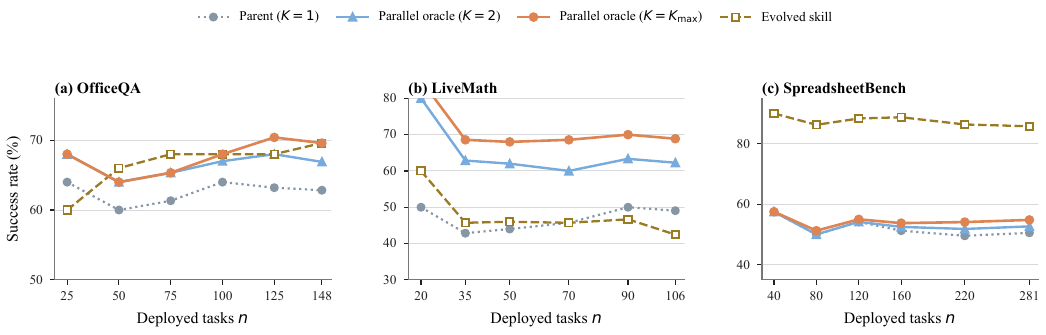}
\caption{Deployment-scale reaggregation of GPT-5.5 TTS outputs under a fixed, outcome-independent SHA-256 item order. $K_{\max}$ is 6, 8, and 4 for OfficeQA, LiveMath, and SpreadsheetBench, respectively. These curves isolate deployment behavior; Figure~\ref{fig:searchqa-tts-amortization} separately gives the SearchQA evolution-cost comparison.}
\label{fig:tts-deployment-scale}
\end{figure}

\begin{table}[H]
\centering
\footnotesize
\setlength{\tabcolsep}{1.0pt}
\renewcommand{\arraystretch}{1.12}
\begin{tabularx}{0.99\textwidth}{@{}L{0.19\textwidth} L{0.12\textwidth} *{9}{C}@{}}
\toprule
\textbf{Benchmark (panel)} & \textbf{Method} & \textbf{Evolved} & $\boldsymbol{K=1}$ & $\boldsymbol{K=2}$ & $\boldsymbol{K=3}$ & $\boldsymbol{K=4}$ & $\boldsymbol{K=5}$ & $\boldsymbol{K=6}$ & $\boldsymbol{K=7}$ & $\boldsymbol{K=8}$ \\
\midrule
\multirow{2}{*}{SearchQA (common 548)} & Parallel oracle & \multirow{2}{*}{\shortstack{417\\(76.09)}} & \shortstack{410\\(74.82)} & \shortstack{419\\(76.46)} & \shortstack{421\\(76.82)} & \shortstack{421\\(76.82)} & \shortstack{422\\(77.01)} & \shortstack{423\\(77.19)} & -- & -- \\
& Sequential last & & \shortstack{410\\(74.82)} & \shortstack{411\\(75.00)} & \shortstack{410\\(74.82)} & \shortstack{411\\(75.00)} & \shortstack{410\\(74.82)} & \shortstack{411\\(75.00)} & -- & -- \\
\midrule
\multirow{2}{*}{OfficeQA (148)} & Parallel oracle & \multirow{2}{*}{\shortstack{103\\(69.59)}} & \shortstack{93\\(62.84)} & \shortstack{99\\(66.89)} & \shortstack{102\\(68.92)} & \shortstack{103\\(69.59)} & \shortstack{103\\(69.59)} & \shortstack{103\\(69.59)} & -- & -- \\
& Sequential last & & \shortstack{93\\(62.84)} & \shortstack{95\\(64.19)} & \shortstack{94\\(63.51)} & \shortstack{92\\(62.16)} & \shortstack{94\\(63.51)} & \shortstack{92\\(62.16)} & -- & -- \\
\midrule
\multirow{2}{*}{DocVQA (374)} & Parallel oracle & \multirow{2}{*}{\shortstack{343\\(91.71)}} & \shortstack{344\\(91.98)} & \shortstack{347\\(92.78)} & \shortstack{350\\(93.58)} & \shortstack{352\\(94.12)} & -- & -- & -- & -- \\
& Sequential last & & \shortstack{344\\(91.98)} & \shortstack{344\\(91.98)} & \shortstack{343\\(91.71)} & \shortstack{342\\(91.44)} & -- & -- & -- & -- \\
\midrule
\multirow{2}{*}{LiveMath (106)} & Parallel oracle & \multirow{2}{*}{\shortstack{45\\(42.45)}} & \shortstack{52\\(49.06)} & \shortstack{66\\(62.26)} & \shortstack{71\\(66.98)} & \shortstack{71\\(66.98)} & \shortstack{72\\(67.92)} & \shortstack{73\\(68.87)} & \shortstack{73\\(68.87)} & \shortstack{73\\(68.87)} \\
& Sequential last & & \shortstack{52\\(49.06)} & \shortstack{56\\(52.83)} & \shortstack{59\\(55.66)} & \shortstack{53\\(50.00)} & \shortstack{51\\(48.11)} & \shortstack{47\\(44.34)} & \shortstack{44\\(41.51)} & \shortstack{51\\(48.11)} \\
\midrule
\multirow{2}{*}{SpreadsheetBench (281)} & Parallel oracle & \multirow{2}{*}{\shortstack{241\\(85.77)}} & \shortstack{142\\(50.53)} & \shortstack{148\\(52.67)} & \shortstack{150\\(53.38)} & \shortstack{154\\(54.80)} & -- & -- & -- & -- \\
& Sequential last & & \shortstack{142\\(50.53)} & \shortstack{138\\(49.11)} & \shortstack{134\\(47.69)} & \shortstack{127\\(45.20)} & -- & -- & -- & -- \\
\bottomrule
\end{tabularx}

\vspace{0.25em}
\begin{tabularx}{0.99\textwidth}{L{0.28\textwidth} C C C C}
\toprule
\textbf{SearchQA full dynamic panel} & \textbf{Parent $S_{\mathrm{parent}}$} & \textbf{Evolved $S_{\mathrm{evo}}$} & $\boldsymbol{S_{\mathrm{par}}^{\mathrm{dyn}}}$ & $\boldsymbol{S_{\mathrm{seq}}^{\mathrm{dyn}}}$ \\
\midrule
Successes (score, $n=1{,}400$) & 1,059 (75.64) & 1,091 (77.93) & 1,085 (77.50) & 1,061 (75.79) \\
Change from parent (points) & 0.00 & $+2.29$ & $+1.86$ & $+0.14$ \\
\bottomrule
\end{tabularx}
\normalsize
\caption{Complete GPT-5.5 test-time-scaling results. Entries report successes and hard-score percentages. SearchQA common support contains 548 items receiving all six attempts; the dynamic panel aggregates item-specific budgets over all 1,400 items. SpreadsheetBench uses the selected Fail-only skill. Each evolved skill receives one call per item; DocVQA compares independent parent executions.}
\label{tab:scaling-numeric}
\end{table}

\subsection{Output-Locked Verifier Sensitivity}

We rescore outcome-blind 100-item panels of GPT-5.5 parent and evolved-skill outputs with the original verifier $V_0$ and a separately fixed $V_1$. Let $\Delta_{V_j}=\operatorname{Score}_{V_j}(s_{\mathrm{evo}})-\operatorname{Score}_{V_j}(s_{\mathrm{parent}})$. With outputs fixed, $\Delta_{V_1}-\Delta_{V_0}$ isolates differential rescoring. The verifiers disagree on 48/1,000 verdicts (4.8\%), while the measured gain is unchanged for SpreadsheetBench, DocVQA, and LiveMath, changes by 1.0 point for OfficeQA, and changes from $-3.0$ to $0.0$ for SearchQA. This output-locked comparison quantifies differential rescoring under one separately fixed alternative verifier per benchmark.

\begin{table}[H]
\centering
\footnotesize
\renewcommand{\arraystretch}{1.12}
\textbf{Output-locked verifier sensitivity}\\[2pt]
\begin{tabularx}{0.62\textwidth}{L{0.36\textwidth} C C C}
\toprule
\textbf{Benchmark} & $\boldsymbol{\Delta_{V_0}}$ & $\boldsymbol{\Delta_{V_1}}$ & \textbf{Diff.} \\
\midrule
SearchQA & $-3.0$ & $0.0$ & $+3.0$ \\
SpreadsheetBench (N) & $+32.0$ & $+32.0$ & $0.0$ \\
OfficeQA & $+5.0$ & $+6.0$ & $+1.0$ \\
DocVQA & $0.0$ & $0.0$ & $0.0$ \\
LiveMath & $-8.0$ & $-8.0$ & $0.0$ \\
\bottomrule
\end{tabularx}
\vspace{0.35em}

\begin{tabularx}{0.99\textwidth}{@{}L{0.14\textwidth} L{0.30\textwidth} L{0.39\textwidth} Y@{}}
\toprule
\textbf{Benchmark} & $\boldsymbol{V_0}$ & $\boldsymbol{V_1}$ & $\boldsymbol{V_1}$ \textbf{identifier} \\
\midrule
SearchQA &
Normalized exact match against gold aliases &
Blind GPT-5.5 semantic-equivalence judge with a fixed JSON verdict schema; three judgments per output &
Prompt SHA \texttt{9ec4149d} \\
OfficeQA &
Adapter-normalized exact match &
Zero-call decimal-exact comparison for numeric answers and canonical normalized-text equality otherwise &
Verifier SHA \texttt{15e15cad} \\
SpreadsheetBench &
Benchmark workbook test cases &
Zero-call, \texttt{data\_only} strict type-and-value equality on every required answer cell &
Verifier SHA \texttt{15e15cad} \\
LiveMath &
Parsed option-label exact match &
Zero-call exact option label or uniquely matched normalized option text &
Verifier SHA \texttt{15e15cad} \\
DocVQA &
ANLS success at threshold 0.999 &
Zero-call membership in the set of punctuation-stripped, case-folded, normalized gold aliases &
Verifier SHA \texttt{15e15cad} \\
\bottomrule
\end{tabularx}
\normalsize
\caption{Output-locked GPT-5.5 verifier sensitivity and verifier definitions. The upper table reports $\Delta_{V_0}$, $\Delta_{V_1}$, and their difference on 100 common non-abstain items per benchmark; the lower table specifies both verdict functions and the alternative-verifier identifier. SpreadsheetBench uses the Normal branch in this diagnostic; N denotes Normal. Scores are percentages and differences are percentage points.}
\label{tab:verifier-numeric}
\end{table}

\end{document}